\documentclass[conference]{IEEEtran}

\def\istechreport{}
\usepackage{cite}

\usepackage{amsmath, amsthm}
\usepackage{amsfonts, amssymb}
    \theoremstyle{plain}
    \newtheorem{lemma}{Lemma}[section]
    \newtheorem{theorem}[lemma]{Theorem}
    \newtheorem*{theorem*}{Theorem}

    \theoremstyle{definition}
    
    \newtheorem{definition}[lemma]{Definition}

    \theoremstyle{plain}

\usepackage{thmtools, thm-restate}

\usepackage{algorithmic}
\usepackage{graphicx}
\usepackage{subcaption}
\usepackage{textcomp}
\usepackage[dvipsnames]{xcolor}
\usepackage{booktabs}
\usepackage{mathtools}
\usepackage[binary-units]{siunitx}

\usepackage{paralist}

\usepackage{url}
\usepackage{hyperref}
\usepackage{verbatim}
\usepackage{listings}
\usepackage{lstautogobble}
\usepackage{tikz}
    \usetikzlibrary{
        arrows,
        arrows.meta,
        backgrounds,
        calc,
        chains,
        decorations,
        decorations.pathreplacing,
        decorations.pathmorphing,
        decorations.markings,
        fit,
        matrix,
        patterns,
        positioning,
        shadows,
        shapes}
\usepackage{pgfplots}
\usepackage{pgfplotstable}
\pgfplotsset{compat=1.14}
\usepgfplotslibrary{units}

\usepackage{balance}

\usepackage{xifthen}
\usepackage{xparse}
\usepackage{etoolbox}

\usepackage{xspace}
\newcommand\ie{i.\,e.\xspace}

\newcommand\eg{e.\,g.\xspace}

\newcommand\etal{et~al.\xspace}
\newcommand\wrt{w.\,r.\,t.\xspace}

\newcommand\LLVMTA{\textsc{llvmta}\xspace}

\NewDocumentEnvironment{techreport}{}
{ 
    \ifthenelse{\isundefined{\istechreport}}%
    {%
        \comment%
    }{%
    }%
}{ 
    \ifthenelse{\isundefined{\istechreport}}%
    {%
    }{%
    }%
}

\NewDocumentEnvironment{paperonly}{}
{
    \ifthenelse{\isundefined{\istechreport}}%
    {%
    }{%
        \comment%
    }%
}{
    \ifthenelse{\isundefined{\istechreport}}%
    {%
    }{%
    }%
}

\usepackage[firstpage]{draftwatermark}
\SetWatermarkText{\hspace*{180mm}\raisebox{224mm}{\includegraphics[width=24mm]{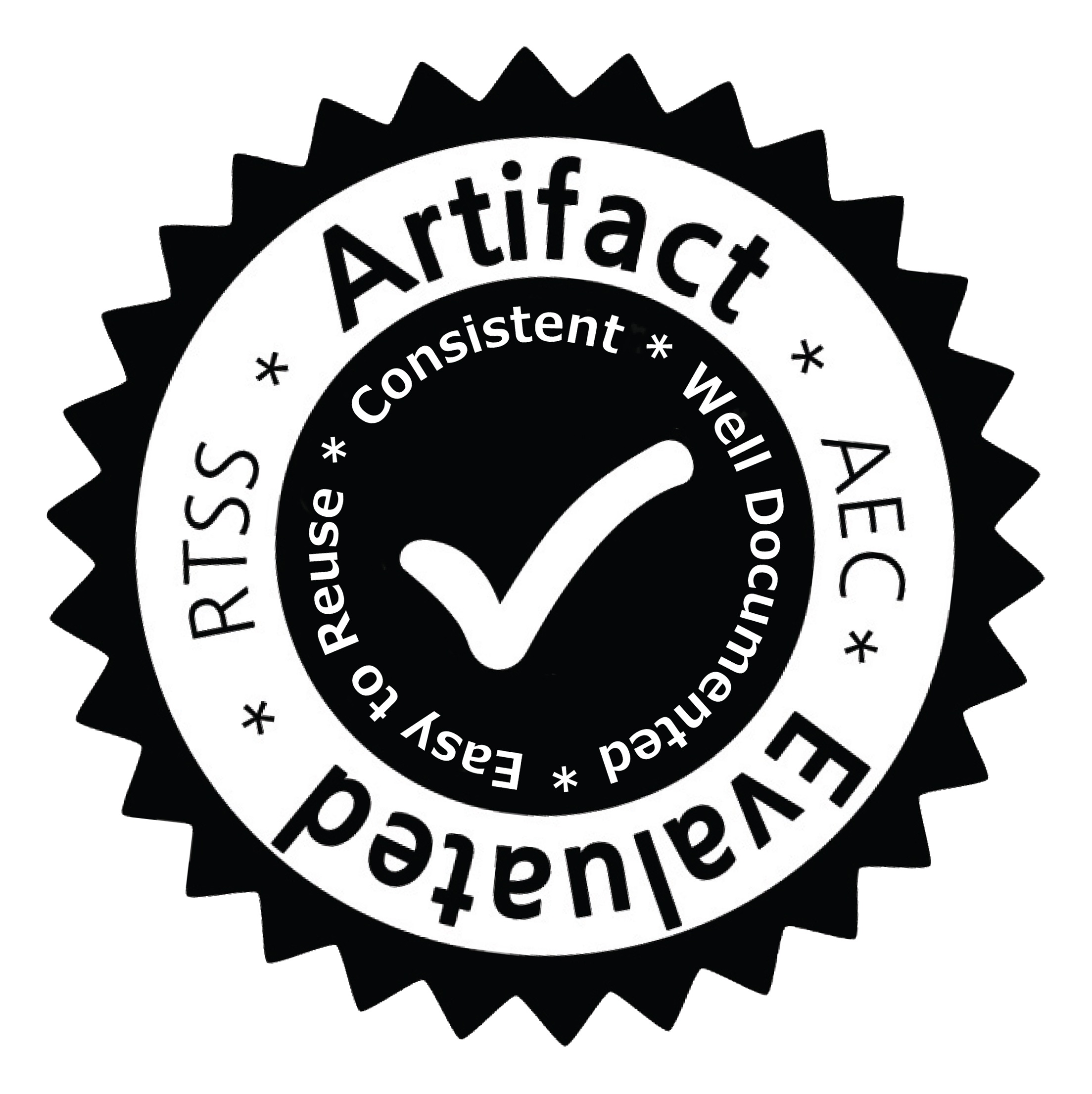}}}
\SetWatermarkAngle{0}

\newcommand{\mi}[1]{\ensuremath{\mathit{#1}}\xspace}
\newcommand{\semantics}[2][]{\left\llbracket#2\ifthenelse{\isempty{#1}}{\right\rrbracket}{\right\rrbracket_{#1}}}
\newcommand{\NN}{\mathbb{N}}

\newcommand{\BB}{\mathbb{B}}

\newcommand{\powerset}[1]{\mathcal{P}(#1)}

\providecommand\where{} 
\newcommand\SetSymbol[1][]{%
    \nonscript\;#1\vert
    \allowbreak
    \nonscript\;\mathopen{}
}
\DeclarePairedDelimiterX\curly[1]{\lbrace}{\rbrace}{%
    \renewcommand\where{\SetSymbol[\delimsize]}#1
}
\DeclarePairedDelimiterX\abs[1]{\lvert}{\rvert}{%
    \ifblank{#1}{\:\cdot\:}{#1}
}
\DeclarePairedDelimiterX\norm[1]{\lVert}{\rVert}{%
    \ifblank{#1}{\:\cdot\:}{#1}
}

\DeclarePairedDelimiter\round{\lparen}{\rparen}
\DeclarePairedDelimiter\brackets{\lbrack}{\rbrack}

\newcommand{\set}{\curly}

\newcommand{\bcs}{\mi{Block\text{-}CS}}
\newcommand{\gcs}{\mi{Global\text{-}CS}}
\newcommand{\must}{\mi{Must}}
\newcommand{\cmust}{\mi{C\text{-}Must}}
\newcommand{\cmmuele}{\cmust{}\times\must{}\times\bcs{}}

\newcommand{\cmay}{\mi{C\text{-}May}}

\newcommand{\xxecsb}{0}
\newcommand{\xxecsm}{\uparrow}
\newcommand{\xxecsc}{\leq k}

\newcommand{\ecs}{\mi{Exact\text{-}CS}}
\newcommand{\ecsb}{\mi{Exact\text{-}CS_{\xxecsb}}}
\newcommand{\ecsm}{\mi{Exact\text{-}CS_{\xxecsm}}}
\newcommand{\ecsc}{\mi{Exact\text{-}CS_{\xxecsc}}}

\newcommand{\blocks}{\mathcal{B}}

\newcommand{\lam}[2]{\lambda #1.\ #2}

\newcommand{\classify}[2][]{%
    \ifthenelse{\isempty{#2}}%
    {\widehat{\mi{persistent}}_{#1}}
    {\widehat{\mi{persistent}}_{#1}\round{#2}}
}

\newcommand{\update}[2][]{%
    \ifthenelse{\isempty{#2}}%
    {\widehat{\mi{update}}_{#1}}
    {\widehat{\mi{update}}_{#1}\round{#2}}
}

\newcommand{\potupdate}[2][]{%
    \ifthenelse{\isempty{#2}}%
    {\widehat{\mi{potentialUpdate}}_{#1}}
    {\widehat{\mi{potentialUpdate}}_{#1}\round{#2}}
}

\newcommand{\abstraces}[1][]{\widehat{C}_{#1}}

\newcommand{\concret}[2][]{%
    \ifthenelse{\isempty{#2}}%
    {\gamma_{#1}}
    {\gamma_{#1}\round{#2}}
}
\newcommand{\abstraction}[2][]{%
    \ifthenelse{\isempty{#2}}%
    {\alpha_{#1}}
    {\alpha_{#1}\round{#2}}
}

\newcommand{\initial}[1][\ecs]{\widehat{\mathcal{I}}_{#1}}

\newcommand{\join}[1][\ecs]{\sqcup_{#1}}

\tikzset{%
    n/.style={draw, circle, thick, inner sep=1mm, minimum width=7mm},
    l/.style={draw, rounded corners=1mm, thick, inner sep=1mm},
    ll/.style={draw, rounded corners=1mm, shape=rectangle split, rectangle split parts=2, thick, inner sep=1mm, font=\vphantom{Q}}, 
    e/.style={shorten >=1mm, shorten <=1mm, thick, ->, >=stealth},
    i/.style={initial, initial text={}},
    every initial by arrow/.style={thick, ->, >=stealth}
}

\begin{document}

\newcommand{\expers}{\textsc{EXPERS}\xspace}
\newcommand\todo[1]{\textcolor{red}{[Todo: #1]}}
\renewcommand\todo[1]{}

\newcommand{\myvspace}[1]{\vspace{#1}}
\renewcommand{\myvspace}[1]{}

\title{Cache Persistence Analysis: Finally Exact}

\author{\IEEEauthorblockN{Gregory Stock, Sebastian Hahn, and Jan Reineke}
\IEEEauthorblockA{\textit{Saarland University}\\
{Saarland Informatics Campus}\\
Saarbr\"ucken, Germany \\
\{g.stock, sebastian.hahn, reineke\}@cs.uni-saarland.de}
}

\maketitle

\begin{abstract}
    Cache persistence analysis is an important part of worst-case execution time (WCET) analysis.
    It has been extensively studied in the past twenty years.
    Despite these efforts, all existing persistence analyses are approximative in the sense that they are not guaranteed to find all persistent memory blocks.

    In this paper, we close this gap by introducing the first exact persistence analysis for caches with least-recently-used (LRU) replacement.
    To this end, we first introduce an exact abstraction that exploits monotonicity properties of LRU to significantly reduce the information the analysis needs to maintain for exact persistence classifications.
    We show how to efficiently implement this abstraction using zero-suppressed binary decision diagrams (ZDDs) and introduce novel techniques to deal with uncertainty that arises during the analysis of data caches.

    The experimental evaluation demonstrates that the new exact analysis is competitive with state-of-the-art inexact analyses in terms of both memory consumption and analysis run time, which is somewhat surprising 
    as we show that persistence analysis is NP-complete.
    We also observe that while prior analyses are not exact in theory they come close to being exact in practice.
\end{abstract}

\myvspace{3cm}


\section{Introduction}\label{sec:introduction}

    Modern processors can perform several arithmetic and logic operations in a single cycle.
    On the other hand, a single access to main memory can take hundreds of cycles.
    To bridge this performance gap, modern processors include one or multiple levels of caches.
    Caches are small but fast memories that store parts of main memory to quickly serve accesses to commonly used instructions and data.
    Memory accesses that ``hit'' the cache are served from the cache at a low latency, while accesses that ``miss'' the cache are served from main memory at a much higher latency.
    The execution time of a program thus heavily depends on how effective the processor's caches are in hiding the high latency of main memory.

    Real-time systems are systems that, in order to function correctly, have to perform their computations with limited amounts of wall-clock time.
    To verify a system's real-time behavior, a major task is to bound each software component's worst-case execution time (WCET).
    In the presence of caches, WCET analysis~\cite{Wilhelm2008} has to account for the software's cache behavior.
    Simply assuming that each memory access could result in a cache miss would yield extremely pessimistic WCET bounds.
    Thus, static cache analyses~\cite{Lv2016} have been developed to soundly and precisely characterize a program's cache behavior on a particular cache architecture.
    These can broadly be categorized into two groups:
    \begin{enumerate}
        \item \emph{Classifying cache analyses} aim to classify individual memory accesses as cache hits or cache misses.
        \item \emph{Quantitative cache analyses} aim to determine the number of cache misses resulting from a set of memory accesses.
    \end{enumerate}
    In this paper we study \emph{persistence analysis}, an instance of quantitative cache analysis.
    Persistence analysis considers all memory accesses in a program, or a fragment of a program such as a loop, that access the \emph{same} memory block.
    A memory block is persistent if all memory accesses referring to this memory block may cumulatively result in {at most one} cache miss during any possible program execution.

    \begin{figure}
        \centering%
        \begin{minipage}[c]{0.4\linewidth}
            \begin{tikzpicture}[node distance=15mm]
                \node[fill, circle, minimum size=2.5mm, inner sep=0pt] (S0) {};
                \node[fill, circle, minimum size=2.5mm, inner sep=0pt, right=of S0] (S1) {};
                \draw[e] (S0) edge[bend right=50] node[draw, thick, rectangle, rounded corners, inner sep=0pt, minimum size=5mm, fill=white] {$y$} (S1);
                \draw[e] (S0) edge[bend left=50] node[draw, thick, rectangle, rounded corners, inner sep=0pt, minimum size=5mm, fill=white] {$x$} (S1);
                \draw[e] (S1) -- ($(S1)+(0, 0.9)$) -| (S0);
                \draw[e] (S1) -- ($(S1)+(0.75, 0)$);
                \draw[e] ($(S0)+(-0.75, 0)$) -- (S0);
            \end{tikzpicture}
        \end{minipage}%
        \begin{minipage}[c]{0.55\linewidth}
            \caption{Simple motivating example for persistence analysis.\label{fig:motivatingexample}}
        \end{minipage}%
    \end{figure}
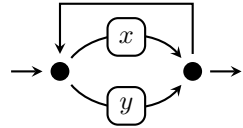

    For a motivating example, consider Figure~\ref{fig:motivatingexample}, which contains the control-flow graph of a simple program.
    The program consists of a loop, in which, in each loop iteration either memory block $x$ or memory block $y$ is accessed.
    As neither block $x$ nor block $y$ is guaranteed to have been accessed in any loop iteration, it is impossible for a classifying cache analysis to classify any of the memory accesses in the program as a guaranteed cache hit, and so a WCET analysis would have to pessimistically account for misses upon all memory accesses.
    However, provided the cache is large enough to hold blocks~$x$ and $y$ simultaneously, among all memory accesses to $x$ (and similarly to $y$) only the very first may result in a cache miss.
    Both $x$ and $y$ are \emph{persistent} and WCET analysis can safely account for at most two misses in total.

    Given a program, the goal of persistence analysis is to determine which of the memory blocks accessed in the program are persistent.
    Persistence analysis has been extensively studied for caches with least-recently-used (LRU) replacement, starting with Mueller's~\cite{Mueller94,Arnold94,White97,Mueller2000rts} and Ferdinand's~\cite{Ferdinand97,Ferdinand1999rts} work in the 1990s up until today~\cite{Ballabriga2008, Cullmann2011, Huynh2011, Nagar2012, Nagar2012thesis, Cullmann2013thesis,Cullmann2013tecs,Zhang15, Reineke2018b}.
    Notably, all prior persistence analyses are \emph{approximative}, in the sense that they are not guaranteed to find all persistent memory blocks of a program.

    In this paper, we close this gap by introducing the first \emph{exact} persistence analysis.
    We develop this analysis via a sequence of three consecutive exact abstractions.
    The first abstraction is based on the observation that the persistence of a memory block can be determined by examining its possible \emph{conflict sets}, \ie, the sets of blocks that may have been accessed since the last access to the block itself.
    The two following abstractions exploit a monotonicity property of LRU replacement to further increase analysis efficiency without sacrificing exactness.

    Next, we discuss how to efficiently implement the exact abstraction using zero-suppressed binary decision diagrams (ZDDs), a data structure that enables the compact representation of sets of conflict sets, sharing information across program points and between different memory blocks.
    We also introduce novel techniques to deal with uncertainty about the memory-access behavior that arises in data cache analysis.

    We experimentally evaluate the new exact persistence analysis and a selection of prior persistence analyses and make the following high-level observations:
    \begin{itemize}
        \item Our exact analysis is competitive with prior analyses in terms of both memory consumption and analysis run time.
        \item While prior persistence analyses are not exact in theory, they come very close to being exact in practice.
    \end{itemize}
    Even though our exact persistence analysis is fairly efficient in practice, its worst-case complexity is exponential.
    We show that persistence analysis is NP-complete, which implies that a persistence analysis that is polynomial in all input parameters is not attainable, unless P=NP.

\section{Background: Caches, Control-Flow Graphs, and Cache Persistence Analysis}\label{sec:background}

\subsection{Caches}

    Caches are fast but small memories that buffer parts of the large but slow main memory in order to bridge the speed gap between the processor and main memory.
    Caches operate at the granularity of memory blocks $b \in \blocks$, which are stored in the cache in \emph{cache lines} of the same size.
    In order to facilitate the cache lookup, the cache is organized in \emph{sets} such that each memory block maps to a unique cache set.
    The size~$k$ of a cache set is called the \emph{associativity} of the cache.
    If an accessed block resides in the cache, the access \emph{hits} the cache.
    Upon a cache \emph{miss}, the block is loaded from main memory.
    To ease the formal presentation in this paper, we assume a fully-associative cache, \ie, with a single cache set.
    Set-associative caches with $n$~sets can be treated as $n$~independent fully-associative caches as described in~\cite{Alt1996}.

    Upon a cache miss, another memory block has to be evicted due to the limited size of the cache.
    The block to evict is determined by the \emph{replacement policy}.
    In this paper, we assume the least-recently-used (LRU) policy that replaces the block that has not been accessed for the longest.
    A memory block~$b$ hits in an LRU~cache of associativity~$k$ if $b$ has been accessed before and less than $k$~distinct blocks have been accessed since the last access to~$b$.
    LRU is generally considered to be the most predictable replacement policy~\cite{Reineke2007}.

    In this paper, we refer to the \emph{age} of block~$b$ as the number of distinct blocks since the last access to~$b$ including the access to $b$ itself\footnote{This is subtly different from most of the related work in which the age of a block does not account for the access of the block itself.}.
    Thus, a block~$b$ hits the cache if its age is less than or equal to the associativity~$k$.

\subsection{Programs as Control-Flow Graphs}

    In this paper, we follow the common approach of representing the program under analysis by its control-flow graph~(CFG).
    A CFG $\mathcal{G}=\round{V, E, i}$ consists of a set of vertices~$V$, corresponding to control locations in the program; a set of edges $E\subseteq V\times \blocks \times V$, which represent the possible control flow between locations; and the initial control location $i \in V$.
    Each edge is annotated with a single memory block accessed between the source and target location.

    The CFG is an abstraction of the program behavior as it does not capture the functional semantics of the instructions.
    In particular, all paths in the graph are assumed to be feasible even if, in reality, some are not, \eg, in case of a nested conditional statement with contradicting conditions.
    All our claims of exactness are relative to this control-flow graph abstraction.
    Incorporating the program semantics into the persistence analysis problem immediately renders it undecidable due to Rice's theorem.
    For example, it is undecidable whether a certain path in a program is feasible, and hence, it is also undecidable to collect all memory access sequences which are needed for exact persistence analysis.

    To ease the visualization of a CFG, we allow empty edges on which no access is performed.
    For the sake of simplicity, we do not treat such empty edges in the formalization, but the extension would be trivial as such empty edges do not influence the cache state.

\subsection{Notion of Persistence}

    A memory block is \emph{persistent} during a program's execution if all accesses to the memory block collectively result in at most one cache miss.
    Assuming the cache is empty at the start of the program's execution, the first access to any memory block will always result in a cache miss.
    Thus, to be persistent, all accesses but the very first to a memory block must hit the cache.

    In other words, a memory block is \emph{persistent} during a program's execution if all accesses to the block hit the cache if the block has been accessed before.
    As an example, in Figure~\ref{fig:motivatingexample}, as discussed before, blocks~$x$ and~$y$ are persistent in a cache of associativity two.

    Due to its dependence on previous accesses, persistence is a property of execution traces.
    A \emph{trace} $\tau = b_0 b_1 \ldots b_{n-1} \in \blocks^*$ is a sequence of memory blocks.
    We use $\tau_i$ to denote the $i$-th block~$b_i$ in the sequence and $\abs{\tau} \coloneqq n$ to denote its length.
    The trace of length zero is denoted by $\epsilon$.

    The \emph{conflict set} of block~$b$ on trace~$\tau$ is the set of all memory blocks accessed from the last access of block~$b$ onward:
    \begin{align*}
        \mi{CS}\round{\tau, b} \coloneqq \bigcap_{\substack{0\leq i<\abs{\tau} \\ \tau_i = b}} \set{\tau_j \where j \geq i}
    \end{align*}
    If block~$b$ has not been accessed on trace~$\tau$, the empty intersection yields the universe~$\blocks$.

    The cardinality of the conflict set is referred to as the \emph{age} of block~$b$ at the end of trace~$\tau$:
    \begin{align*}
        \mi{age}\round{\tau, b} \coloneqq \abs{\mi{CS}\round{\tau, b}}
    \end{align*}

    In accordance with the discussion at the beginning of this subsection, we call block~$b$ persistent on trace~$\tau$ if $b$ is cached once it has been accessed before.
    \begin{definition}[Persistence on Trace]\label{def:persistence-trace}
        Memory block $b\in \blocks$ is \emph{persistent on trace}~$\tau \in \blocks^*$ if:
        \begin{align*}
            \round{\exists 0 \leq i < \abs{\tau} : \tau_i = b} \rightarrow \mi{age}\round{\tau, b} \leq k
        \end{align*}
    \end{definition}

    The above definition captures persistence of a block on a single trace.
    In order to reason about the persistence of a memory block during a program's execution, we need to capture all possible traces of a given program.
    To this end, we define the \emph{trace semantics}, which captures for each location in a control-flow graph all possible traces that end in this location.

    Formally, the trace semantics is defined as the least solution~$R^C$ of the following set of equations where $R^C\colon V\to \powerset{\blocks^*}$ maps each location to a set of traces:
    \begin{align*}
        \forall w\in V: R^C\round{w} = \mathcal{I}^C\round{w} \cup \bigcup_{\round{v, b, w}\in E} \mi{update}\round*{R^C\round{v}, b}
    \end{align*}
    where $\mi{update}\round{T, b}$ is defined as $\set{\tau \circ b \where \tau \in T}$ and $\circ$ denotes the concatenation operator.
    The initial values, denoted by~$\mathcal{I}^C\round{w}$, are the empty set for all locations except the program's entry point which is initialized to the set containing the empty trace~$\mathcal{I}^C\round{i} = \set{\epsilon}$.

    The equations can intuitively be understood as follows:
    Either a trace starts in location $w$, then it is given by $\mathcal{I}^C\round{w} = \set{\epsilon}$; or it starts elsewhere and reaches $w$ from one of its predecessors~$v$ via edge $\round{v, b, w}$.
    In the latter case, the trace reaching location $w$ is obtained by concatenating the memory access~$b$ on the edge from $v$ to $w$ to the trace reaching location~$v$.

    A memory block~$b$ is persistent throughout a given program if $b$ is persistent on each trace that may be immediately followed by an access to~$b$.
    The set $V_b = \set{v \in V \where \exists w \in V: \round{v, b, w} \in E}$ contains the program locations that can be followed by an access to block~$b$.
    \begin{definition}[Persistence]\label{def:persistence-program}
        Memory block $b\in \blocks$ is \emph{persistent}, denoted by $\mi{persistent}\round{b}$, if at each location $v\in V_b \subseteq V$ that can be followed by an access to~$b$:
        \begin{align*}
            \mi{persistent}\round*{R^C\round{v}, b} \coloneqq \forall \tau \in R^C\round{v} : \mi{persistent}\round{\tau, b}
        \end{align*}
    \end{definition}
    A memory block~$b$ may cause at most one miss during any possible execution through $\mathcal{G}$ if and only if it is persistent according to the definition above. 

    \paragraph*{Scopes}
    A memory block might only persist in the cache during a certain phase of execution, \eg, in the innermost loop of a loop nest, and not during the overall program execution.
    To capture this behavior, \emph{scopes} have been introduced in~\cite{Ballabriga2008,Huynh2011} to describe a portion of program execution.
    A block that is persistent within a scope can cause at most one miss for each entrance of the scope during execution.
    For the sake of readability, we limit our formalization to persistence within the whole program.
    It is an easy exercise to extend the definitions to account for scopes.
    Indeed, the experimental evaluation in Section~\ref{sec:expeval} is performed by analyses at scope level.

\subsection{Existing Cache Persistence Analyses}\label{sec:existing}

    In general, the  trace semantics is not computable as the number of traces may be infinite and the lengths of individual traces are unbounded.
    Persistence analyses thus rely on abstractions of cache traces that lead to finite representations.

    A variety of cache persistence analyses has been proposed in the literature~\cite{Arnold94,Mueller94,White97,Ferdinand97,Ferdinand1999rts,Mueller2000rts,Ballabriga2008,Cullmann2011,Huynh2011,Nagar2012,Nagar2012thesis,Cullmann2013tecs,Cullmann2013thesis,Zhang15} with varying degrees of precision.
    In this section, we briefly discuss two of the existing persistence analyses, $\cmust$ and $\bcs$ to convey how such persistence analyses operate in general, and also to illustrate that they are not exact.

    Observe that according to Definition~\ref{def:persistence-program}, whether or not a memory block~$b$ is persistent is determined by the sizes of $b$'s conflict sets at all program points that may be followed by accesses to block~$b$.
    As a consequence, all existing persistence analyses can be seen as \emph{approximating} the possible sets of conflict sets of each memory block at each program point.

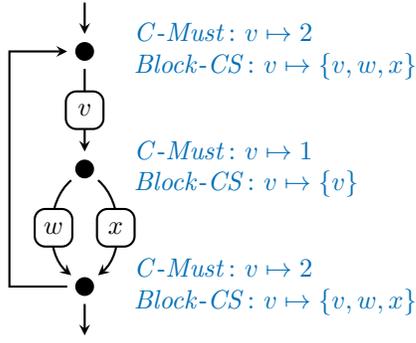
\begin{figure}
        \centering%
        \begin{tikzpicture}[node distance=15mm]
            \node[fill, circle, minimum size=2.5mm, inner sep=0pt] (S0) {};
            \node[fill, circle, minimum size=2.5mm, inner sep=0pt, below=of S0, yshift=2mm] (S1) {};
            \node[fill, circle, minimum size=2.5mm, inner sep=0pt, below=of S1, yshift=2mm] (S2) {};
            \draw[e] (S0) edge node[draw, thick, rectangle, rounded corners, inner sep=0pt, minimum size=5mm, fill=white] {$v$} (S1);
            \draw[e] (S1) edge[bend right=50] node[draw, thick, rectangle, rounded corners, inner sep=0pt, minimum size=5mm, fill=white] {$w$} (S2);
            \draw[e] (S1) edge[bend left=50] node[draw, thick, rectangle, rounded corners, inner sep=0pt, minimum size=5mm, fill=white] {$x$} (S2);
            \draw[e] ($(S0)+(0,0.75)$) -- (S0);
            \draw[e] (S2) -- ($(S2)+(-1,0)$) |- (S0);
            \draw[e] (S2) -- ($(S2)+(0,-0.75)$);
            \node[right=4mm of S0, align=left, NavyBlue] {$\cmust{}\colon v\mapsto 2$ \\ $\bcs{}\colon v\mapsto \set{v, w, x}$};
            \node[right=4mm of S1, align=left, NavyBlue] {$\cmust{}\colon v\mapsto 1$ \\ $\bcs{}\colon v\mapsto \set{v}$};
            \node[right=4mm of S2, align=left, NavyBlue] {$\cmust{}\colon v\mapsto 2$ \\ $\bcs{}\colon v\mapsto \set{v, w, x}$};
        \end{tikzpicture}
        \caption{Example illustrating $\cmust$ and $\bcs$, which shows that $\cmust$ may be more precise than $\bcs$.}
        \label{fig:cmust}
        \myvspace{-4mm}
    \end{figure}

    Existing persistence analyses employ two different approaches to approximate the conflict sets of a memory block:
    \begin{enumerate}
        \item The \cmust analysis~\cite{Ferdinand97,Ferdinand1999rts} maintains an \emph{upper bound on the sizes} of all of the block's possible conflict sets.
        \item The \bcs analysis~\cite{Huynh2011,Cullmann2013thesis} maintains a \emph{superset} of all of the block's possible conflict sets.
    \end{enumerate}
    See Figure~\ref{fig:cmust} for an example illustrating the results of the two analyses.
    For readability, the figure only includes the analysis information for memory block~$v$.
    Following the access to $v$, the only possible conflict set of $v$ is $\set{v}$ and so $\cmust$ maintains a bound of $1$ and \bcs maintains $\set{v}$ as a superset of all of $v$'s conflict sets.
    After the possible accesses to~$w$ and~$x$, the possible conflict sets of $v$ are $\set{v, w}$ and $\set{v, x}$, and $\cmust\round{v} = 2$.
    However, to overapproximate both conflict sets, $\bcs\round{v} = \set{v, w, x} = \set{v, w} \cup \set{v, x}$.
    Thus, $\cmust$ is able to conclude that $v$ is persistent in a cache of size~$2$, while \bcs is not, as $\abs{\set{v, w, x}} > 2$.
    Clearly, $\bcs$ is not an exact analysis.

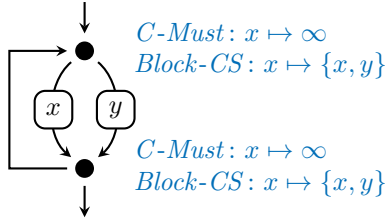
\begin{figure}
        \centering%
        \begin{tikzpicture}[node distance=15mm]
            \node[fill, circle, minimum size=2.5mm, inner sep=0pt] (S0) {};
            \node[fill, circle, minimum size=2.5mm, inner sep=0pt, below=of S0, yshift=2mm] (S1) {};
            \draw[e] (S0) edge[bend right=50] node[draw, thick, rectangle, rounded corners, inner sep=0pt, minimum size=5mm, fill=white] {$x$} (S1);
            \draw[e] (S0) edge[bend left=50] node[draw, thick, rectangle, rounded corners, inner sep=0pt, minimum size=5mm, fill=white] {$y$} (S1);
            \draw[e] (S1) -- ($(S1)+(-1,0)$) |- (S0);
            \draw[e] (S1) -- ($(S1)+(0,-0.75)$);
            \draw[e] ($(S0)+(0,0.75)$) -- (S0);
            \node[right=4mm of S0, align=left, NavyBlue] {$\cmust{}\colon x\mapsto \infty$ \\ $\bcs{}\colon x\mapsto \set{x, y}$};
            \node[right=4mm of S1, align=left, NavyBlue] {$\cmust{}\colon x\mapsto \infty$ \\ $\bcs{}\colon x\mapsto \set{x, y}$};
        \end{tikzpicture}
        \caption{Example illustrating \cmust and \bcs, which shows that \bcs may be more precise than \cmust.}
        \label{fig:blockcs}
        \myvspace{-4mm}
    \end{figure}

    Unfortunately, as the example in Figure~\ref{fig:blockcs} illustrates, $\cmust$ neither is exact:
    Both $x$ and $y$ are persistent in a cache of size $2$, and indeed \bcs is able to show that, as $\bcs\round{x} = \set{x, y}$.
    On the other hand, $\cmust$ is not able to derive any finite bound on the sizes of $x$'s possible conflict sets.
    This is because $\cmust$ does not ``remember'' whether or not a given memory block has already been accounted for in its upper bounds.
    This may lead the analysis to account for the same block multiple times in loops.

\subsection{A General Framework for Cache Persistence Analyses}

    Persistence analyses can be formalized within the framework of abstract interpretation~\cite{Cousot77} to reason about their correctness and precision.
    Here, we briefly present a simplified version of the persistence analysis framework developed in~\cite{Reineke2018b}.
    Persistence abstractions~$\abstraces{}$ are characterized by
    \begin{itemize}
        \item an abstract update function $\update{}$ to model the effect of a memory access,
        \item a join operator $\join[]{}$ to combine multiple abstract traces into one at control-flow joins, and
        \item an abstract function $\classify{}$ to classify memory blocks as persistent.
    \end{itemize}
    Note that the join operator also defines a partial order $\sqsubseteq$ on the abstract traces as follows: $x \sqsubseteq y$ if and only if $x \join[]{} y = y$.
    This partial order captures the relative precision of different abstract traces, where $x \sqsubseteq y$ implies that $x$ is more precise analysis information than $y$.

    In order to formally capture the meaning of abstract traces, abstraction and concretization functions, $\abstraction{}$ and $\concret{}$, can be defined to relate sets of concrete traces to abstract traces.

    Analogously to the concrete semantics, the abstract semantics is captured as the least solution~$\widehat{R}\colon V \rightarrow \abstraces{}$ of the following set of equations:
    \begin{align*}
        \forall w\in V : \widehat{R}\round{w} = \initial[]{}\round{w} \sqcup \bigsqcup_{\round{v, b, w}\in E} \update{}\round[\big]{\widehat{R}\round{v}, b}
    \end{align*}
    The initial value~$\initial[]{}\round{w}$ is $\bot$, the bottom element of the partial order $\sqsubseteq$, for all locations except for the program's entry point, where $\initial[]{}\round{i} = \abstraction{}\round*{\mathcal{I}^C\round{i}}$.

    A sound persistence analysis overapproximates the concrete trace semantics, \ie, $R^C\round{v} \subseteq \concret{}\round[\big]{\widehat{R}\round{v}}$ for all locations~$v$.
    Equivalently, given that $\round{\abstraction{}, \concret{}}$ form a Galois connection~\cite{Cousot77}, we have $\abstraction{}(R^C\round{v}) \sqsubseteq \widehat{R}\round{v}$ for all locations~$v$.

    Analogously to persistence in the concrete case, a memory block~$b$ is classified as persistent in a given program, denoted by $\classify{}\round{b}$, if $b$ is classified as persistent at each control location in $V_b$, \ie, each location that might be immediately followed by an access to~$b$.
    A sound persistence analysis never classifies a block as persistent that is not actually persistent on all concrete traces:

    \begin{definition}[Soundness of Persistence Analysis]
        A persistence analysis is \emph{sound} if:
        \begin{align*}
            \classify{}\round{b} \rightarrow \mi{persistent}\round{b}
        \end{align*}
    \end{definition}

    Sound abstractions are not guaranteed to be exact, \ie, there can be persistent memory blocks that are \emph{not} classified as persistent by the abstraction.
    Indeed, none of the existing persistence analyses is exact~\cite{Reineke2018b}.

\section{Exact Cache Persistence Analysis:\texorpdfstring{\\}{ }A Sequence of Abstractions}\label{sec:exactanalysis}

    In this paper, we do not just aim for yet another sound persistence analysis, but we aim for an \emph{exact} persistence analysis that determines each and every persistent memory block.

    \begin{definition}[Exactness of Persistence Analysis]\label{def:exact}
        A persistence analysis is \emph{exact} if:
        \begin{align*}
            \classify{}\round{b} \leftrightarrow \mi{persistent}\round{b}
        \end{align*}
    \end{definition}
    Note that by definition an exact analysis is also sound.
    How can we obtain an exact analysis?
    This requires an abstraction that satisfies the following two properties:
    \begin{enumerate}
        \item
            Applying the abstraction to ``perfect'' concrete information preserves enough information to precisely classify memory blocks as persistent or not.
            This property is comparably easy to achieve.
            It is captured formally by~(\ref{con:classify}) in the theorem below.
            The $\cmust$ abstraction, for example, satisfies this property, while $\bcs$ does not.
        \item
            Abstract joins and abstract updates may not lose any additional information, beyond the information loss inherent to the abstraction itself.
            This is more difficult to achieve, and indeed none of the existing persistence analyses does.
            This property is captured formally by (\ref{con:transfer}) and (\ref{con:abstraction}) in the theorem below.
    \end{enumerate}

    \begin{restatable}[Exactness of Persistence Analysis]{restheorem}{thmexact}\label{thm:exact}%
        A persistence analysis over a finite abstract domain~$\abstraces$ is \emph{exact} if:
        \begin{align}
            \forall T, b &: \abstraction{\mi{update}\round{T, b}} = \update{\abstraction{T}, b} \label{con:transfer} \\
            \forall I, T_i &: \abstraction{}\round[\Big]{\bigcup_{i \in I} T_i} = \bigsqcup_{i \in I} \abstraction{T_i} \label{con:abstraction}
        \end{align}
        and the abstraction preserves the persistence classification:
        \begin{align}
            \forall T, b : \mi{persistent}\round{T, b} \leftrightarrow \classify{\abstraction[]{T}, b}\label{con:classify}
        \end{align}
    \end{restatable}
    \begin{proof}
        \newcommand{\trans}{\ensuremath{f}\xspace}
        \newcommand{\abstrans}{\ensuremath{\hat{f}}\xspace}
        \newcommand{\lfp}{\mathit{lfp}\xspace}
        We will use standard arguments from abstract interpretation and fixpoint theory (along the lines of~\cite{Cousot79}) to show that (\ref{con:transfer}) and (\ref{con:abstraction}) imply that:
        \begin{align}
            \forall v\in V : \abstraction{}\round*{R^C\round{v}} = \widehat{R}\round{v} \label{eq:complete-alpha}
        \end{align}
        \ie, the abstract semantics is precisely the abstraction of the concrete semantics.
        Applying (\ref{con:classify}) to (\ref{eq:complete-alpha}) then yields the theorem.

        To this end, we first define concrete and abstract transformers~\trans and \abstrans as follows:
        \begin{align*}
            \trans(R)              &\coloneqq \lam{w}{\mathcal{I}^C\round{w} \cup \bigcup_{\round{v, b, w}\in E} \mi{update}\round{R\round{v}, b}} \\
            \abstrans(\widehat{R}) &\coloneqq \lam{w}{\initial[]{} \round{w} \sqcup \bigsqcup_{\round{v, b, w}\in E} \update{}\round[\big]{\widehat{R}\round{v}, b}}
        \end{align*}
        By construction, the trace semantics~$R^C$ and the abstract semantics~$\widehat{R}$ are the least fixed points of $\trans$ and $\abstrans$.

        Applying (\ref{con:transfer}) and (\ref{con:abstraction}) we show below that:
        \begin{align}
            \alpha \circ \trans = \abstrans \circ \alpha\label{eq:exactabstraction}
        \end{align}
        where $\alpha\colon C \to \abstraces{}$ is lifted to the required domain $\alpha \colon \round{V\to C} \to \round[\big]{V\to \abstraces{}}$ as follows:
        \begin{align}
            \alpha\round{h} \coloneqq \lam{v\in V}{\alpha\round{h\round{v}}} \label{eq:alpha-lift}
        \end{align}
        The transformations to show (\ref{eq:exactabstraction}) are as follows:
        \begin{align*}
            \displaystyle
            \begin{array}{c@{}>{\displaystyle}l}
                & \round{\alpha \circ \trans}\round{R} \\
                \overset{\mi{Def}}{=}{}& \alpha \round{\lam{w}{\mathcal{I}^C\round{w} \cup \bigcup_{\round{v, b, w}\in E} \mi{update}\round{R\round{v}, b}}} \\
                \overset{\round{\ref{eq:alpha-lift}}}{=}{}& \lam{w}{\alpha \round{\mathcal{I}^C\round{w} \cup \bigcup_{\round{v, b, w}\in E} \mi{update}\round{R\round{v}, b}}} \\
                \overset{\round{\ref{con:transfer}}, \round{\ref{con:abstraction}}}{=}{}& \lam{w}{\initial[]{} \round{w} \sqcup \bigsqcup_{\round{v, b, w}\in E} \update{\alpha\round{R\round{v}}, b}} \\
                \overset{\round{\ref{eq:alpha-lift}}}{=}{}& \lam{w}{\initial[]{} \round{w} \sqcup \bigsqcup_{\round{v, b, w}\in E} \update{\alpha\round{R}\round{v}, b}} \\
                \overset{\mi{Def}}{=}{}& \round[\big]{\abstrans \circ \alpha}\round{R}
            \end{array}
        \end{align*}
        By construction, \trans is continuous~\cite{Davey02}, and \abstrans is continuous as it is monotone and the abstract domain is finite.\\
        By Kleene's fixpoint theorem ($\star$) and (\ref{eq:exactabstraction}), we have: 
        \begin{equation}\label{eq:complete-alpha-proof}
            \begin{aligned}
                \abstraction{}\round[\big]{R^C} & \overset{\mi{Def}}= \abstraction{\lfp \trans} \overset{\round{\star}}= \abstraction{}\round[\Big]{\bigcup_{i \in \NN_0} \trans^i\round{\bot}} \\
                & \overset{\round{\ref{con:abstraction}}}= \bigsqcup_{i \in \NN_0} \abstraction{}\round[\big]{\trans^i\round{\bot}} \overset{\round{\ref{eq:exactabstraction}}}= \bigsqcup_{i \in \NN_0} \abstrans^i\round{\abstraction{\bot}} \\
                &= \bigsqcup_{i \in \NN_0} \abstrans^i\round[\big]{\widehat{\bot}} \overset{\round{*}}= \lfp \abstrans \overset{\mi{Def}}= \widehat{R}
            \end{aligned}
        \end{equation}
        Applying (\ref{con:classify}) to (\ref{eq:complete-alpha-proof}) then yields the theorem.
    \end{proof}

\begin{figure}
        \centering%
        \begin{tikzpicture}
            \node (A) {$\powerset{\blocks^*}$};
            \node[right=13mm of A] (B) {$\abstraces[\xxecsb]{}$};
            \node[right=15mm of B] (C) {$\abstraces[\xxecsm]{}$};
            \node[right=15mm of C] (D) {$\abstraces[\xxecsc]{}$};
            \draw[e] (A) edge[bend left] node[above] {$\abstraction[\xxecsb]{}$} (B);
            \draw[e] (B) edge[bend left] node[below] {$\concret[\xxecsb]{}$} (A);
            \draw[e] (B) edge[bend left] node[above] {$\abstraction[\xxecsm]{}$} (C);
            \draw[e] (C) edge[bend left] node[below] {$\concret[\xxecsm]{}$} (B);
            \draw[e] (C) edge[bend left] node[above] {$\abstraction[\xxecsc]{}$} (D);
            \draw[e] (D) edge[bend left] node[below] {$\concret[\xxecsc]{}$} (C);
        \end{tikzpicture}
        \myvspace{-1mm}
        \caption{Exact persistence abstractions developed in this section.}
        \label{fig:overview-trace-abstractions}
        \myvspace{-3mm}
    \end{figure}
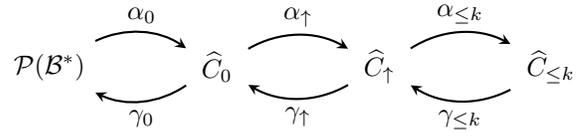

    The remaining part of this section presents the abstraction of cache traces underlying our new exact persistence analysis.
    For pedagogical reasons, the abstraction is presented incrementally in three steps:
    First, an abstraction of concrete traces as a mapping from memory blocks to sets of conflict sets is defined.
    In the second step, a monotonicity property of LRU replacement is exploited to reduce the number of conflict sets the analysis needs to track.
    Third, conflict sets that exceed the cache's associativity are collapsed to further improve efficiency.
    The overall scheme is depicted in Figure~\ref{fig:overview-trace-abstractions}.

\subsection{From Memory-Access Traces to Sets of Conflict Sets}

    The first abstraction $\ecsb$ maintains for each memory block the set of all possible \emph{conflict sets} that appear in cache traces in which the memory block has been accessed.
    Recall that the conflict set of block~$b$ is the set of all distinct memory blocks accessed since the last access of~$b$.
    Formally, the abstract domain is defined as:
    \begin{align*}
        \abstraces[\xxecsb] \coloneqq \blocks \to \powerset{\powerset{\blocks}}
    \end{align*}
    An abstract trace $\widehat{S}\in\abstraces[\xxecsb]$ represents all concrete traces where for each memory block $b\in\blocks$ either $b$'s conflict set is in~$\widehat{S}\round{b}$ or $b$ has not been accessed on the trace.

    Upon a memory access to block~$b$, its set of conflict sets is set to $\set{\set{b}}$, \ie, $b$ conflicts just with itself.
    For all remaining memory blocks~$b'$, the accessed block~$b$ is added to \emph{every} conflict set $s\in\widehat{S}\round{b'}$:
    \begin{equation}\label{eq:ecsb-update}
        \begin{multlined}[c][.85\displaywidth]
            \update[\xxecsb]{}\round[\big]{\widehat{S}, b} \coloneqq \\
            \lambda b'.
            \begin{cases}
                \set{\set{b}}                                               & \text{if $b' = b$} \\
                \set[\big]{s \cup \set{b} \where s\in\widehat{S}\round{b'}} & \text{otherwise}
            \end{cases}
        \end{multlined}
    \end{equation}
    Note that as long as a block~$b'$ has not been accessed, the information $\widehat{S}\round{b'} = \emptyset$ propagates.

    At control-flow joins in the program, the union of the sets of conflict sets is taken:
    \begin{align*}
        \forall I\subseteq\NN_0 : \sideset{}{_\xxecsb} \bigsqcup\limits_{i\in I} \widehat{S}_i \coloneqq \lam{b\in\blocks}{\bigcup_{i\in I} \widehat{S}_i(b)}
    \end{align*}
    A drawback of the $\bcs$ abstraction is the loss of precision at joins as seen in Figure~\ref{fig:cmust}.
    Using sets of sets, $\ecsb$ keeps all conflict sets side by side without losing any precision.

    Next, we define the abstraction function~$\abstraction[\xxecsb]{}$ that relates a set of concrete traces to an abstract trace.
    For singleton sets, \ie, sets containing a single concrete trace, the abstraction function is recursively defined as follows:
    \begin{align*}
        \abstraction[\xxecsb]{\set{\epsilon}} &\coloneqq \lam{b \in \blocks}{\emptyset} \\
        \abstraction[\xxecsb]{\set{\tau \circ b_{n-1}}} &\coloneqq \update[\xxecsb]{\abstraction[\xxecsb]{\set{\tau}}, b_{n-1}}
    \end{align*}
    Abstractions of concrete traces are recursively constructed using the abstract update function with the accessed memory blocks.
    The base case, \ie, the abstraction of the empty trace~$\epsilon$, assigns an empty set of conflict sets to each block~$b$ since no memory block has been accessed.

    The abstraction~$\abstraction[\xxecsb]{}$ is lifted to arbitrary sets in the usual way:\looseness=-1
    \begin{align*}
        \abstraction[\xxecsb]{\set{t_i \where i \in I\subseteq \NN_0}} \coloneqq \sideset{}{_\xxecsb}\bigsqcup_{i\in I} \abstraction[\xxecsb]{\set{t_i}}
    \end{align*}

    A memory block~$b$ is classified as persistent if all conflict sets have cardinality less than or equal to the associativity~$k$.
    This means that at most~$k-1$ distinct other blocks have been accessed since the last access to the block itself:
    \begin{align*}
        \classify[\xxecsb]{}\round[\big]{\widehat{S}, b} \coloneqq \max\set[\big]{\abs{s} \where s\in \widehat{S}\round{b}} \leq k
    \end{align*}
    Note that we assume the maximum over an empty set to be zero.
    As a consequence, a block that has never been accessed is classified as persistent.

\begin{figure}
        \centering%
        \begin{tikzpicture}[node distance=15mm]
            \node[fill, circle, minimum size=2.5mm, inner sep=0pt, label=below:{$l_0$}] (S0) {};
            \node[fill, circle, minimum size=2.5mm, inner sep=0pt, label=below:{$l_1$}, right=of S0] (S1) {};
            \node[fill, circle, minimum size=2.5mm, inner sep=0pt, label=below:{$l_2$}, right=of S1, xshift=-5mm] (S15) {};
            \node[fill, circle, minimum size=2.5mm, inner sep=0pt, label=below:{$l_3$}, right=of S15] (S2) {};
            \node[fill, circle, minimum size=2.5mm, inner sep=0pt, label=below:{$l_4$}, right=of S2] (S3) {};
            \draw[e] (S0) edge node[draw, thick, rectangle, rounded corners, inner sep=0pt, minimum size=5mm, fill=white] {$v$} (S1);
            \draw[e] (S1) edge (S15);
            \draw[e] (S15) edge[bend left=40] node[draw, thick, rectangle, rounded corners, inner sep=0pt, minimum size=5mm, fill=white] {$w$} (S2);
            \draw[e] (S15) edge[bend right=40] node[draw, thick, rectangle, rounded corners, inner sep=0pt, minimum size=5mm, fill=white] {$x$} (S2);
            \draw[e] (S2) edge[bend left=40] node[draw, thick, rectangle, rounded corners, inner sep=0pt, minimum size=5mm, fill=white] {$y$} (S3);
            \draw[e] (S2) edge[bend right=40] (S3);
            \draw[e] (S2) |- ($(S2)+(0,0.8)$) -| (S15);
            \draw[e] ($(S0)-(0.75,0)$) -- (S0);
            \draw[e] (S3) -- ($(S3)+(0.75,0)$);
        \end{tikzpicture}
        \caption{Example illustrating $\ecs$.}
        \label{fig:ecs-example}
    \end{figure}
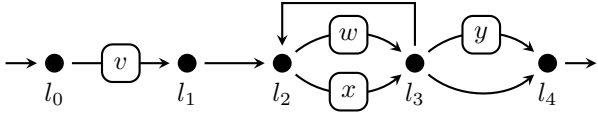

    See Figure~\ref{fig:ecs-example} for an example of a control-flow graph and the corresponding $\ecs$ cache trace abstractions in Table~\ref{tab:ecs-example}.
    Note that only the analysis information for memory block~$v$ is shown as it highlights best the differences between the different exact analyses.

    \begin{restatable}[Exactness of $\ecsb$]{restheorem}{thmexactecsb}\label{thm:exact-ecsb}
        The persistence analysis~$\ecsb$ is exact in the sense of Definition~\ref{def:exact}.
    \end{restatable}

    Due to space limitations we omit the proofs in this paper, but they can be found in \ifthenelse{\isundefined\istechreport}{a technical report~\cite{Stock2019}}{the appendix of this technical report}.

    \begin{table*}
        \caption{Concrete traces as regular expressions and persistence abstractions for Figure~\ref{fig:ecs-example} (with $k = 3$).}
        \label{tab:ecs-example}
        \myvspace{-1mm}
        \centering%
        \def\tablename{Table}%
        \begin{tabular}{cclll}
            \toprule
                  & Reachable access traces                   & $\ecsb$                                                                                    & $\ecsm$                       & $\ecsc$                     \\\midrule
            $l_0$ & $\varepsilon$                             & $v\mapsto\emptyset$                                                                        & $v\mapsto\emptyset$           & $v\mapsto\emptyset$         \\
            $l_1$ & $v$                                       & $v\mapsto\set{\set{v}}$                                                                    & $v\mapsto\set{\set{v}}$       & $v\mapsto\set{\set{v}}$     \\
            $l_2$ & $v\brackets{w|x}^*$                       & $v\mapsto\set{\set{v}, \set{v,w}, \set{v,x}, \set{v,w,x}}$                                 & $v\mapsto\set{\set{v,w,x}}$   & $v\mapsto\set{\set{v,w,x}}$ \\
            $l_3$ & $v\brackets{w|x}^+$                       & $v\mapsto\set{\set{v,w}, \set{v,x}, \set{v,w,x}}$                                          & $v\mapsto\set{\set{v,w,x}}$   & $v\mapsto\set{\set{v,w,x}}$ \\
            $l_4$ & $v\brackets{w|x}^+$, $v\brackets{w|x}^+y$ & $v\mapsto\set{\set{v,w}, \set{v,x}, \set{v,w,x}, \set{v,w,y}, \set{v,x,y}, \set{v,w,x,y}}$ & $v\mapsto\set{\set{v,w,x,y}}$ & $v\mapsto\set{\blocks}$     \\\bottomrule
        \end{tabular}
        \myvspace{-2mm}
    \end{table*}

\subsection{Exploiting Monotonicity}\label{sec:exploiting-monotonicity}

    While \ecsb is exact, the number of conflict sets to track for a given block can be high.
    To classify a block as persistent only the largest conflict set of each block at each program location is relevant.
    It would be tempting to only keep the largest sets, but, unfortunately, this would yield an incorrect analysis, as these largest sets could not be correctly maintained across updates and joins.

    It is, however, safe to remove conflict sets that are completely subsumed by others.
    This is because the update function~$\update[\xxecsb]{}$ is \emph{monotonic}:
    For example, consider the set of conflict sets $\set{\set{a, b, c}, \set{a, b}}$.
    No matter which blocks are accessed and used to update these conflict sets, the first set $\set{a, b, c}$ and its successors will always subsume the second set $\set{a, b}$ and its successors.
    Removing such subsumed sets reduces the computational effort of the analysis, in particular the memory consumption, without any loss of precision.
    The resulting abstraction $\ecsm$ is defined relative to the $\ecsb$ abstraction as shown in Figure~\ref{fig:overview-trace-abstractions}.

    The abstraction function $\abstraction[\xxecsm]{}$ takes an abstract trace and removes all conflict sets that are subsumed by larger sets:\looseness=-1
    \begin{align*}
        \abstraction[\xxecsm]{}\round[\big]{\widehat{S}} &\coloneqq \lam{b\in\blocks}{\mi{maxSet}\round[\big]{\widehat{S}\round{b}}}
    \end{align*}
    where $\mi{maxSet}$ is defined as follows:
    \begin{align*}
        \mi{maxSet}\round{S} \coloneqq \set{s\in S \where \neg\exists s'\in S : s\subsetneq s'}
    \end{align*}

    In order to maintain a minimal set of conflict sets, the join operator takes the maximal conflict sets of the union of the sets of conflict sets to be joined:
    \begin{align}\label{eq:ecsm-join}
        \forall I\subseteq\NN_0 : \sideset{}{_\xxecsm} \bigsqcup\limits_{i\in I} \widehat{S}_i \coloneqq \lam{b\in\blocks}{\mi{maxSet}\round[\Big]{\bigcup\limits_{i\in I} \widehat{S}_i\round{b}}}
    \end{align}
    Similarly, the abstract update function removes all non-maximal conflict sets after adding~$b$ to the conflict sets:
    \begin{align*}
        \update[\xxecsm]{}\round[\big]{\widehat{S}, b} \coloneqq \lam{b'\in\blocks}{}\mi{maxSet}\round[\big]{\update[\xxecsb]{}\round[\big]{\widehat{S}, b}\round{b'}}
    \end{align*}
    The classification function can be reused without modification.

    \begin{restatable}[Exactness of $\ecsm$]{restheorem}{thmexactecsm}\label{thm:exact-ecsm}
        The persistence analysis~$\ecsm$ is exact in the sense of Definition~\ref{def:exact}.
    \end{restatable}

\subsection{Limit at Associativity}

    The abstraction described above allows the individual conflict sets to grow arbitrarily large.
    However, the persistence classification checks the existence of a single conflict set with cardinality greater than the associativity~$k$.
    Thus, there is no need to distinguish conflict sets containing more than $k$ elements.\looseness=-1

    Instead, all conflict sets with a cardinality larger than $k$ can be collapsed into a single representative~$\blocks$, \ie, the set of all memory blocks.
    Note that all conflict sets are trivially subsumed by~$\blocks$.
    This fact and the monotonicity property from Section~\ref{sec:exploiting-monotonicity} allow to replace the whole set of conflict sets by $\set{\blocks}$ in the presence of an oversized conflict set.
    This reduces the computational effort of the analysis even further, in particular its memory consumption, without any loss of precision.
    The resulting abstraction $\ecsc$ is defined relative to the $\ecsm$ abstraction as shown in Figure~\ref{fig:overview-trace-abstractions}.

    The abstraction function $\abstraction[\xxecsc]{}$ takes an $\ecsm$ abstract trace and eliminates conflict sets with cardinality larger than~$k$:
    \begin{align*}
        \abstraction[\xxecsc]{}\round[\big]{\widehat{S}} \coloneqq \lam{b\in\blocks}{\mi{limit}\round[\big]{\widehat{S}\round{b}}}
    \end{align*}
    where
    \begin{align*}
        \mi{limit}\round{S} \coloneqq \begin{cases}
                                          \set{\blocks} & \text{if $\exists s \in S : \abs{s} > k$} \\
                                          S             & \text{otherwise}
                                      \end{cases}
    \end{align*}
    The update function~$\update[\xxecsc]{}$ is consequently defined as:
    \begin{align*}
        \update[\xxecsc]{}\round[\big]{\widehat{S}, b} \coloneqq \lam{b'\in\blocks}{\mi{limit}\round[\big]{\update[\xxecsm]{}\round[\big]{\widehat{S}, b}\round{b'}}}
    \end{align*}
    The join operator and the classification function can be reused as they do not increase the size of any conflict set.

    \begin{restatable}[Exactness of $\ecsc$]{restheorem}{thmexactecsc}\label{thm:exact-ecsc}
        The persistence analysis~$\ecsc$ is exact in the sense of Definition~\ref{def:exact}.
    \end{restatable}

\newcommand{\mytimes}{{\times}\allowbreak}

\subsection{Example of Superiority over Prior Persistence Analyses}\label{sec:exactisbestexample}

    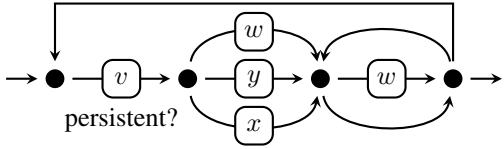
\begin{figure}
        \centering%
        \begin{tikzpicture}[node distance=15mm]
            \node[fill, circle, minimum size=2.5mm, inner sep=0pt] (S0) {};
            \node[fill, circle, minimum size=2.5mm, inner sep=0pt, right=of S0] (S1) {};
            \node[fill, circle, minimum size=2.5mm, inner sep=0pt, right=of S1] (S2) {};
            \node[fill, circle, minimum size=2.5mm, inner sep=0pt, right=of S2] (S3) {};
            \draw[e] (S0) edge node[draw, thick, rectangle, rounded corners, inner sep=0pt, minimum size=5mm, fill=white] (p) {$v$} (S1);
            \draw[e] (S1) edge[bend left=80] node[draw, thick, rectangle, rounded corners, inner sep=0pt, minimum size=5mm, fill=white] {$w$} (S2);
            \draw[e] (S1) edge[bend right=80] node[draw, thick, rectangle, rounded corners, inner sep=0pt, minimum size=5mm, fill=white] {$x$} (S2);
            \draw[e] (S1) edge node[draw, thick, rectangle, rounded corners, inner sep=0pt, minimum size=5mm, fill=white] {$y$} (S2);
            \draw[e] (S2) edge node[draw, thick, rectangle, rounded corners, inner sep=0pt, minimum size=5mm, fill=white] {$w$} (S3);
            \draw[e] (S2) edge[bend right=80] (S3);
            \draw[e] (S3) edge[bend right=80] (S2);
            \draw[e] (S3) |- ($(S3)+(0, 1.0)$) -| (S0);
            \node[below=of p, yshift=1.5cm] {persistent?};
            \draw[e] ($(S0)-(0.75, 0)$) -- (S0);
            \draw[e] (S3) -- ($(S3)+(0.75, 0)$);
        \end{tikzpicture}
        \caption{Example illustrating exact analysis outperforming previous persistence analyses. Fully-associative cache with $k=3$.}
        \label{fig:exact-superior}
    \end{figure}

    Figure~\ref{fig:exact-superior} contains an example control-flow graph on which all existing persistence analysis fail.
    In the example, block~$v$ is clearly persistent in a fully-associative cache of size ${k = 3}$: at most two distinct blocks are accessed between any two accesses to~$v$.

    The most precise known persistence analysis is the combination $\cmust \mytimes\must \mytimes\bcs$ of the traditional must analysis, the $\cmust$, and the $\bcs$ analysis~\cite{Reineke2018b}.

    Since neither $x$, $y$, nor $w$ is \emph{guaranteed} to get accessed, the must analysis does not gain any information about them.
    The \bcs analysis fails to classify $v$ as persistent because there are three distinct blocks that may conflict with $v$ in between two consecutive accesses to~$v$.
    As a consequence, neither of the two analyses is able to support the \cmust analysis.
    On its own, the \cmust analysis ages~$v$ upon each access distinct from~$v$.
    Since more than associativity many accesses might happen between the accesses to~$v$ (note the inner loop), the $\cmust$ analysis is also of no use.

\section{Exact Cache Persistence Analysis:\texorpdfstring{\\}{ }Implementation}\label{sec:implementation}

\subsection{Implementation using Binary Decision Diagrams}

    The implementation of the \ecsc analysis needs to maintain a set of conflict sets for each memory block in the program at every program point.
    Maintaining separate explicit representations of each set of conflict sets for each memory block and at each program point would likely be highly inefficient.
    An efficient implementation should implicitly (and thus hopefully more compactly) represent sets of conflict sets and it should share as much information as possible between different memory blocks and program points.

    Observe that sets of conflict sets can be represented using Boolean functions:
    Let the set of memory blocks be $\blocks = \set{b_1, \dots, b_n}$.
    Then a Boolean valuation $v = v_1, \dots, v_n \in \BB^{n}$ represents a set~$\mi{set}\round{v}$ of memory blocks as follows:
    \begin{align*}
        \mi{set}\round{v_1, \dots, v_n} = \set{b_i \where 1 \leq i \leq n \wedge v_i = 1}
    \end{align*}
    Extending upon this, a Boolean function $f \colon \BB^{n} \rightarrow \BB$ represents a set of conflict sets~$\mi{CS}\round{f}$ as follows:
    \begin{align*}
        \mi{CS}\round{f} \coloneqq \set{\mi{set}\round{v} \where v \in \BB^{n} \wedge f\round{v} = 1}
    \end{align*}

    Binary decision diagrams~\cite{Drechsler2001} are a class of data structures that have been designed to efficiently represent and manipulate Boolean functions.
    They compactly represent individual Boolean functions and share information between the representations of different Boolean functions stored in the same data structure.
    Our implementation uses zero-suppressed binary decision diagrams (ZDDs)~\cite{Minato2001, Mishchenko2001}, a variant of binary decision diagrams optimized to represent sets of sparse sets.
    This is beneficial in our setting as the sets of conflict sets used in our trace abstractions are typically sparse with respect to the universe~$\blocks$; no conflict set in \ecsc is greater than the associativity~$k$.

    To perform operations on ZDDs, the Colorado University Decision Diagram (\texttt{CUDD}) library\footnote{Available at \url{https://github.com/ivmai/cudd}}~\cite{Somenzi2018} is used in combination with the \texttt{EXTRA} library\footnote{Available at \url{https://people.eecs.berkeley.edu/~alanmi/research/extra/}}.
    The latter offers extended ZDD procedures that facilitate the manipulation of ZDDs.
    More precisely, the following operators of the \texttt{EXTRA} library are used in the persistence analysis where $S$ and $T$ are sets containing sets:
    \begin{enumerate}
        \item
            $\mi{maxUnion}$ calculates the maximum of the union of sets~$S$ and $T$:
            \begin{align*}
                \mi{maxUnion}\round{S, T} \coloneqq \mi{maxSet}\round{S\cup T}
            \end{align*}
            This function is used to compute the abstract joins exploiting monotonicity in~(\ref{eq:ecsm-join}).
        \item
            $\mi{maxDotProduct}$ takes all pair-wise unions of subsets from $S$ and $T$ and computes the maximum of this set, \ie, it drops subsumed elements in the result:
            \begin{multline*}
                \mi{maxDotProduct}\round{S, T} \coloneqq\\ \mi{maxSet}\set{s\cup t\where s\in S \wedge t\in T}
            \end{multline*}
            This function is involved in the abstract update~(\ref{eq:ecsb-update}) to add an accessed block~$b$ to every conflict set in a set.
            More precisely, the following property is exploited:
            \begin{align*}
                \mi{maxDotProduct}\round{S, \set{\set{b}}} = \mi{maxSet}\set{s\cup \set{b}\where s\in S}
            \end{align*}
    \end{enumerate}

\subsection{Extension to Data Caches}

    \begin{figure}
        \begin{lstlisting}[language=C, basicstyle=\small\ttfamily, numbersep=1em, xleftmargin=2em]
            int arr[10];
            for (int i = 0; i < 100; ++i)
                sum += arr[read_sensor()]
        \end{lstlisting}
        \caption{Input-dependent data accesses.}
        \label{lst:array-access}
    \end{figure}

    Until now, we have assumed that only a single known memory block is accessed on each control-flow edge.
    This is valid in the context of an instruction cache analysis.
    For data cache analysis, however, we have to deal with uncertainty about which memory blocks might be accessed.
    The memory blocks a single \texttt{load} or \texttt{store} instruction accesses might depend on dynamic aspects, \eg, the program input or a loop iteration counter.
    See Listing~\ref{lst:array-access} for an example of a memory access depending on program input.
    Array~\texttt{arr} is small compared to the number of loop iterations in the example.
    If \texttt{arr} completely fits into the cache, an exact persistence analysis has to classify it as persistent.

    Instead of cache updates with a single accessed memory block, the update function for data caches has to consider a set of potentially accessed memory blocks $B\subseteq \blocks$.
    The straightforward solution is to lift the update function for single blocks to sets of blocks by performing an update for each block~$b\in B$ individually and then joining the results:
    \begin{align}
        \update{}\round[\big]{\widehat{S}, B} \coloneqq \bigsqcup_{b\in B}\update{}\round[\big]{\widehat{S}, b}\label{eq:datacacheupdate}
    \end{align}

    \paragraph*{Implementation}
    Prior to cache analysis, a preprocessing analysis determines a set~$B \subseteq \blocks$ of potentially accessed memory blocks for each memory instruction in the program.
    Such sets can be large, \eg, when a large array is accessed.
    The preprocessing analysis might even fail to derive useful information for some accesses, resorting to $\blocks$.
    In such cases, performing a cache update according to (\ref{eq:datacacheupdate}) can be computationally expensive or infeasible.

    To efficiently cope with large sets~$B \subseteq \blocks$, we implemented a slightly different analysis that maps memory blocks to a \emph{list} of $k$ sets of conflict sets, \ie,
    \begin{align*}
        \abstraces{} \coloneqq \blocks \to \powerset{\powerset{\blocks}} \times \cdots \times \powerset{\powerset{\blocks}}
    \end{align*}
    The list entry at index $i \in \brackets{0, \ldots, k-1}$ corresponds to the set of conflicting sets where each conflict set implicitly contains an additional $i$~distinct unknown anonymous memory blocks.

    Upon classification, these anonymous blocks have to be accounted for by adding them to the conflict set cardinalities:
    \begin{align*}
        \classify[]{}\round[\big]{\widehat{S}, b} \coloneqq \forall i : \max\set[\big]{\abs{s} \where s\in \widehat{S}_i\round{b}} + i \leq k
    \end{align*}

    At updates with a single memory block~$b$, the information for~$b$ is reset to the set with the conflict set $\set{b}$ as first list entry.
    For all other memory blocks, the original update is performed for each entry in the list:
    \begin{equation*}
        \begin{multlined}[c][.95\displaywidth]
            \update{}\round[\big]{\widehat{S}, b} \coloneqq \\
            \lambda b'.
            \begin{cases}
                \brackets[\big]{\set{\set{b}}, \emptyset, \ldots, \emptyset}                               & \text{if $b' = b$} \\
                \brackets[\big]{\update{}\round[\big]{\widehat{S}_i, b}\round{b'} \bigm\vert 0 \leq i < k} & \text{otherwise}
            \end{cases}
        \end{multlined}
    \end{equation*}
    In case of a completely unknown access, all entries are shifted by one position to the right.
    If there has been a non-empty conflict set at the rightmost position, the corresponding block is not guaranteed to be cached any more, which is indicated by $\blocks$ in the first position:
    \begin{equation*}
        \begin{multlined}[c][.95\displaywidth]
            \update{}\round[\big]{\widehat{S}} \coloneqq \\
            \lambda b'.
            \begin{cases}
                \brackets[\big]{\hspace{6pt}\emptyset\hspace{6pt}, \widehat{S}_0\round{b'}, \ldots, \widehat{S}_{k-2}\round{b'}} & \text{if $\widehat{S}_{k-1}\round{b'} = \emptyset$} \\
                \brackets[\big]{\set{\blocks}, \hspace{10pt}\emptyset\hspace{10.5pt}, \ldots, \hspace{16pt}\emptyset\hspace{15.4pt}} & \text{otherwise} \\
            \end{cases}
        \end{multlined}
    \end{equation*}
    For sets~$B\subseteq\blocks$ of potentially accessed blocks larger than the associativity~$k$, the above update can be performed without any loss of precision.
    For smaller sets~$B\subseteq\blocks$, the update is performed as defined by Equation~\ref{eq:datacacheupdate}.

\subsection{Validation}

    The two features described in this section are used for testing purposes and by default turned off during analysis.
    Their intention is to validate our implementation and spot bugs.
    One of the advantages of having an exact analysis is that it can be used to verify the correctness of other analyses to some extent as the exact one will always provide more precise results than the imprecise ones.
    We implemented a flag that allows us to run the exact persistence analysis alongside any other persistence analysis.
    After every operation, \eg, updates or joins, the sets of persistent memory blocks are compared.
    The set of the exact analysis must always include the other set due to its exactness.
    This proves by no means the implementation of the exact analysis to be correct but gives some hint that the analyses are computing reasonable results.

    Another critical aspect of the implementation is the manipulation of ZDDs to represent sets of conflict sets.
    Therefore, the analysis was extended with an explicit representation which uses the sets of the C++ Standard Template Library.
    Again, after each operation, the equality between the ZDDs and explicit set representations is checked to see whether the ZDD library is working as intended.

\subsection{Integration with WCET Analysis}

\begin{figure}
        \begin{minipage}{.48\linewidth}
            \begin{tikzpicture}[state/.style={circle, inner sep=0pt, draw}]
                \tikzset{larrow/.style={>=stealth, ->}}
                \node[state] (s1) {$\mu_1$};
                \node[state, label=left:{}, below of=s1, left of=s1, yshift=1mm] (m) {$\mu_2$};
                \node[state, label=left:{}, below of=s1, right of=s1, yshift=1mm] (h) {$\mu_3$};

                \node[state, below of=m, yshift=1mm] (s2) {$\mu_4$};
                \node[state, below of=s2, left of=s2, yshift=1mm] (m2) {$\mu_5$};
                \node[state, below of=s2, right of=s2, yshift=1mm] (h2) {$\mu_6$};

                \draw[larrow] (s1) -- node[left, yshift=1.5mm] {$b$ miss} (m);
                \draw[larrow] (s1) -- node[right, yshift=1.5mm] {$b$ hit} (h);

                \draw[larrow, dashed] ([yshift=0.7cm] s1.center) -- (s1);
                \draw[larrow, dashed] (h) -- ([yshift=-0.7 cm] h.center);

                \draw[larrow, dashed] (m) -- (s2);
                \draw[larrow] (s2) -- node[left, yshift=1.5mm] {$b$ miss} (m2);
                \draw[larrow] (s2) -- node[right, yshift=1.5mm] {$b$ hit} (h2);

                \draw[larrow, dashed] (m2) -- ([yshift=-0.7cm] m2.center);
                \draw[larrow, dashed] (h2) -- ([yshift=-0.7cm] h2.center);
            \end{tikzpicture}
        \end{minipage}%
        \hfill%
        \begin{minipage}{.48\linewidth}
            Integer Linear Program\\[0.3cm]
            Variables:\\
            $x_{\mu_i\rightarrow\mu_j}$ execution frequency\\ \phantom{$x_{\mu_i\rightarrow\mu_j}$} of edge $\mu_i \rightarrow \mu_j$\\[0.2cm]
            Persistence Constraints:\\[-0.4cm]
            \[
                x_{\mu_1\rightarrow\mu_2} + x_{\mu_4\rightarrow\mu_5} \leq 1
            \]
        \end{minipage}%
        \caption{Abstract execution graph snippet and persistence path analysis constraint.}\label{fig:absexecgraph}
    \end{figure}
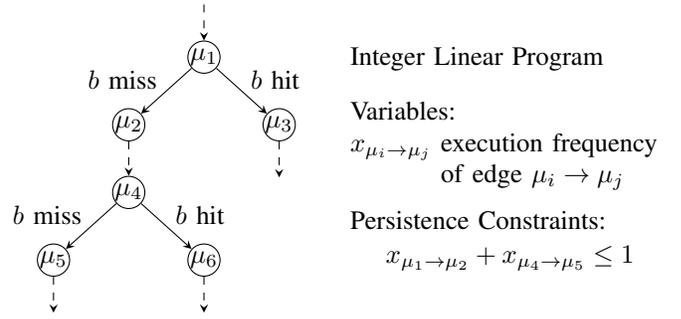

    The information on the persistence of memory blocks is used in worst-case execution time (WCET) analysis to obtain more precise upper bounds on a program's execution time.
    WCET analysis is commonly performed in two phases: microarchitectural analysis and path analysis.
    Microarchitectural analysis constructs an \emph{abstract execution graph} that represents the execution of a program on a given hardware platform at the granularity of processor cycles.
    Nodes correspond to the abstract (microarchitectural) state of the hardware platform, including information about the caches, and edges represent the actual execution.
    Figure~\ref{fig:absexecgraph} illustrates an example graph.

    The path analysis determines the longest path through the abstract execution graph using an integer linear program~\cite{Li1995}.
    Since persistence information is a property of execution traces, it is taken into account in the path analysis.
    To this end, for each persistent memory block $b\in\blocks$ in the program, a linear constraint is used to limit the number of cache misses of $b$ to one.
    An example constraint is shown in Figure~\ref{fig:absexecgraph}.

\section{Related Work}\label{sec:relwork}

    For LRU caches, persistence analysis is strongly related to must analysis~\cite{Ferdinand1999rts}:
    A memory block~$b$ {must be cached} at location~$v$, if $b$'s age is at most $k$, the associativity of the cache, on all traces ending in location~$v$.
    This is the case \emph{if and only if} block~$b$ is persistent on all traces ending in location~$v$ \emph{and} block~$b$ has previously been accessed on all these traces.
    As a consequence, must analysis could be solved by a persistence analysis running alongside a ``definitely-accessed'' analysis that determines whether a block is guaranteed to have been accessed before reaching a given program location.
    Exploring this relation further is future work.

    The development of the exact persistence analyses in this paper has been inspired by recent work of Touzeau~\etal~\cite{Touzeau2019}, in which they develop exact must and may analyses for LRU caches.
    Similarly to our persistence analysis, their implementation employs ZDDs to compactly represent sets of sets of memory blocks and their abstraction exploits the monotonicity of LRU replacement to reduce the number of sets to track.
    Besides solving a related but different problem, our analysis adds support for the efficient analysis of data caches as discussed in Section~\ref{sec:implementation} and unlike~\cite{Touzeau2019} we evaluate the analysis in the context of WCET analysis.

    The presence of timing anomalies~\cite{Lundqvist1999,Reineke2006} can often be traced back to the non-monotonicity of a system's dynamics.
    In contrast to LRU, other cache replacement policies such as FIFO, NMRU, and PLRU, do not behave monotonically, and have been found to exhibit timing anomalies~\cite{Berg2006,Gebhard2010}. 
    For that reason it seems unlikely that our analysis approach can be extended to such policies.
   The strictly in-order core~\cite{Hahn2018} is a pipelined processor core that has been designed to be free of timing anomalies by eliminating dependencies in the pipeline that induce non-monotonicity.

    A variety of cache persistence analyses has been proposed in the literature~\cite{Arnold94,Mueller94,White97,Ferdinand97,Ferdinand1999rts,Mueller2000rts,Ballabriga2008,Cullmann2011,Huynh2011,Nagar2012,Nagar2012thesis,Cullmann2013tecs,Cullmann2013thesis,Zhang15} with varying degrees of precision.
    Many persistence analyses can be seen as combinations of analyses from a small set of basic persistence analyses.
    We have already discussed two of these basic analyses in Section~\ref{sec:background}.: \bcs and \cmust.
    Another basic analysis, called \gcs maintains a superset of the conflict sets of \emph{all} memory blocks, rather than maintaining separate information for each memory block, as \bcs does.
    This simple approach has been particularly popular in the literature~\cite{Arnold94,Mueller94,White97,Mueller2000rts,Cullmann2013tecs,Cullmann2013thesis} and it constitutes the most efficient known analysis.
    
        \newcommand{\versatz}{15mm}
    \newcommand{\versatzxl}{15mm}

    \newcommand{\smallerfont}{\footnotesize}

    \newcommand{\refthm}[1]{}
    \newcommand{\entry}[2]{\smallerfont{}#1}
    \newcommand{\entrytwo}[3]{\smallerfont{}#1}
    \newcommand{\entryW}[3]{\entry{#1}{#2}\\[-1mm]\smallerfont{}#3}
    \newcommand{\entrytwoW}[4]{\entrytwo{#1}{#2}{#3}\\[-1mm]\smallerfont{}#4}

    \begin{figure}
        \begin{center}
            \resizebox{\linewidth}{!}{\begin{tikzpicture}[every text node part/.style={align=center}, node distance = 1.2cm and 0.9cm]
                \node [draw] (CM) at (0,0) {\entryW{\cmust}{cmsoundness}{{\mbox{\cite{Ferdinand97,Ferdinand1999rts}}}}};
                \node [above of=CM, xshift=0*\versatzxl] (CMmay) {\entryW{$\cmust \mytimes\cmay$}{statereductionCMMaysoundness}{\cite{Cullmann2011,Cullmann2013tecs,Cullmann2013thesis,Zhang15}}};
                \node [above of=CM, xshift=2*\versatz] (CMmust) {\entryW{$\cmust \mytimes\must$}{cmmustupdatesoundness}{\cite{Reineke2018b}}};
                \node [above of=CM, xshift=-\versatzxl-\versatzxl,draw] (maycs)  {\entryW{\bcs}{maycssoundness}{\cite{Huynh2011,Cullmann2013thesis}}};
                \node [above of=CMmay, xshift=0*\versatz] (CMmaycs) {\entryW{$\cmust \mytimes\bcs$}{statereductionsoundness}{\cite{Cullmann2013thesis,Zhang15}}};
                \node [above of=CMmust, xshift=0*\versatz] (CMmaymust) {\entrytwoW{$\cmust \mytimes\must \mytimes\cmay$\hspace{0mm}~}{cmmustupdatesoundness}{statereductionCMMaysoundness}{\cite{Nagar2012,Nagar2012thesis}}};
                \node [above of=CMmaymust, draw] (CMmustmaycs) {\entrytwoW{$\cmust \mytimes\must \mytimes\bcs$}{cmmustupdatesoundness}{statereductionsoundness}{\cite{Reineke2018b}}};
                \node [above of=CMmustmaycs, draw] (exactcs) {\entryW{\ecs}{}{[this paper]}};
                \node [below of=maycs] (may) {\entryW{\cmay}{cmaysoundness}{}{\cite{Reineke2018b}}};
                \node [below of=may, draw] (cs) {\entryW{\gcs}{globalmaycssoundness}{\cite{Arnold94,Mueller94,White97,Mueller2000rts,Cullmann2013tecs,Cullmann2013thesis}}};
                \node [below of=CMmust, xshift=0*\versatz] (must) {\entryW{(\must)}{mustsoundness}{\cite{Ferdinand97,Ferdinand1999rts}}};

                \draw [thick] (CM) -- (CMmay);
                \draw [thick] (CM) -- (CMmust);
                \draw [thick] (maycs) -- (CMmaycs);
                \draw [thick] (CMmay) -- (CMmaycs);
                \draw [thick] (CMmay) -- (CMmaymust);
                \draw [thick] (CMmust) -- (CMmaymust);
                \draw [thick] (CMmaymust) -- (CMmustmaycs);
                \draw [thick] (CMmaycs) -- (CMmustmaycs);
                \draw [thick] (CMmustmaycs) -- (exactcs);
                \draw [thick] (cs) -- node[left] {\refthm{cmaymoreprecise}} (may);
                \draw [thick] (may) -- (CMmay);
                \draw [thick] (may) -- node[left] {\refthm{cmaylessprecise}} (maycs);
                \draw [thick] (must) -- (CMmust);

            \end{tikzpicture}}
        \end{center}
        \caption{Landscape of persistence analyses adapted from~\cite{Reineke2018b}.}\label{fig:hassediagram}
    \end{figure}
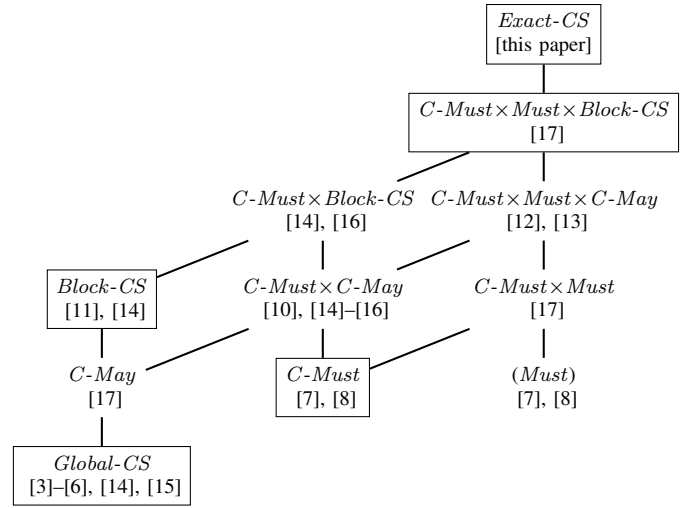

    Reineke~\cite{Reineke2018b} has analyzed the landscape of persistence analyses and has shown how the different analyses relate to each other in terms of precision, and how they can be explained as combinations of basic persistence analyses.
    In Figure~\ref{fig:hassediagram}, we reproduce the landscape of persistence analyses from~\cite{Reineke2018b}.
    In the diagram, a node labeled $A \mytimes B$ denotes an analysis that is obtained by the combination of the basic analyses $A$ and $B$.
    Further, if analysis~$A$ is provably more precise than analysis~$B$, then $A$ and $B$ are connected by an edge and $A$ is higher up in the diagram.

    For the experimental evaluation in Section~\ref{sec:expeval}, we chose to evaluate the three basic analyses \gcs, \bcs, \cmust, and the most precise known combination of analyses $\cmust \mytimes\must \mytimes\bcs$ as explained below in Section~\ref{sec:expeval}.
    These four analyses and the exact analysis developed in this paper are framed in Figure~\ref{fig:hassediagram}.

\section{Experimental Evaluation}\label{sec:expeval}

    In this evaluation, we compare the performance of the exact analysis $\ecs$ in terms of the calculated WCET bound as well as the run time and memory consumption with the four existing persistence analyses $\gcs$, $\bcs$, $\cmust$, and $\cmmuele$.
    While \gcs represents the most efficient persistence analysis, $\cmust\mytimes\must\mytimes\bcs$ provides the most precise results among the previously known analyses.
    In addition, we chose \bcs and \cmust as they represent two basic but complementary ideas to approximate conflict sets as described in Section~\ref{sec:existing}.
    The results below also indicate that a comparison with the various possible other combinations (Figure~\ref{fig:hassediagram}) would not have revealed major further insights.

    \begin{figure}[t]
        \centering%
        \scalebox{0.82}{\includegraphics{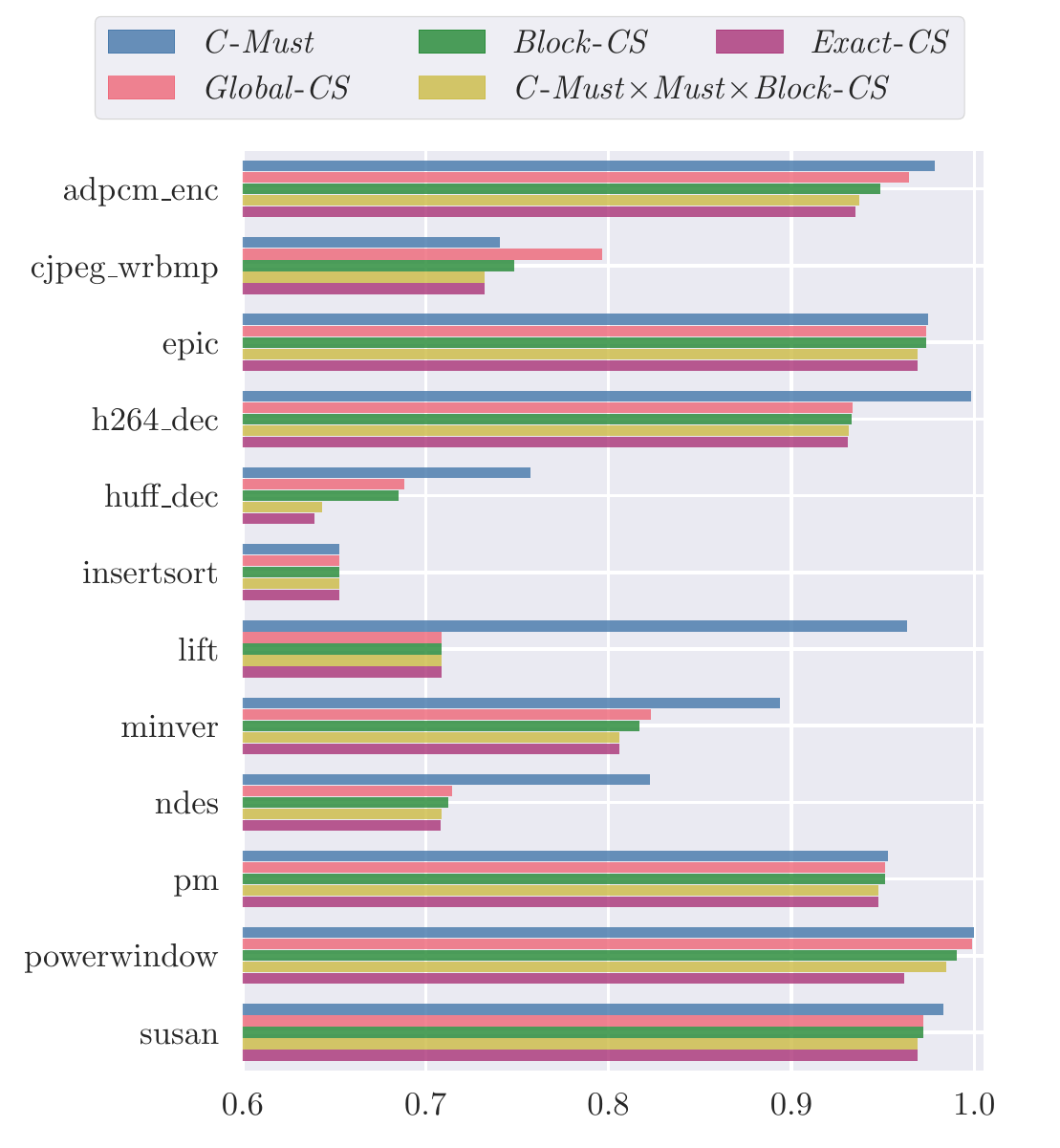}}
        \caption{WCET ratios of the persistence analyses compared with performing no persistence analysis at all.}
        \label{fig:bar-bound}
    \end{figure}

\subsection{Experimental Setup}

    We implemented the exact persistence analysis as well as the previously known analyses within the WCET analysis framework~\LLVMTA{}, which is described in~\cite{Hahn2019}. 
    \LLVMTA{} analyses the persistence of memory blocks at the level of scopes instead of the whole program.
    Precisely, every loop at any nesting level in the program is considered a separate persistence scope.

    To evaluate the implementation of the exact analysis, we use the benchmarks of the TACLeBench suite~\cite{Falk2016}.
    TACLeBench consists of several open-source C programs commonly used to evaluate timing analysis.

    In the following, results and measurements are shown for TACLeBench compiled \emph{without} compiler optimizations enabled.
    Software for safety-critical embedded systems is often compiled without optimizations to ease the subsequent verification of the produced binary \wrt the underlying high-level model~\cite{Franca2011}.
    We also conducted experiments with enabled compiler optimizations.
    Since compiler optimizations generally reduce the amount of memory operations, they are less impacted by different cache persistence analyses and thus show slightly fewer differences among the different analyses.
    For the sake of completeness, all the numbers are made available in \ifthenelse{\isundefined\istechreport}{the technical report~\cite{Stock2019}}{the appendix of this technical report}.

    The analyzed cache configuration has a total size of $\SI{4}{\kilo\byte}$ and consists of $32$ cache sets, $8$ ways, and cache lines holding $16$-byte-sized memory blocks, which is also used by~\cite{Touzeau2019} taking into account the relative small size of the benchmarks.
    Accessing a single word in main memory takes ten processor cycles with an additional cycle per consecutive word accessed.
    In total, the load of a cache line takes 13~cycles which is a realistic value for main memories such as the Automotive DRAM MT46V16M16~\cite{MicronDatasheet} clocked at $\SI{100}{\MHz}$.
    An additional evaluation of the exact analysis with a higher latency of 100~cycles showed no interesting differences.

    All measurements have been performed on the same Linux machine, equipped with an Intel Core{\footnotesize\texttrademark{}} \mbox{i5-4590}~CPU (running at $\SI{3.30}{\GHz}$) and $\SI{20}{\giga\byte}$ of main memory.

\subsection{Analysis Precision}

    This section answers how the exact, ZDD-based analysis compares with inexact alternatives in terms of the calculated WCET bounds.
    Figure~\ref{fig:bar-bound} shows the WCET bounds obtained with the different persistence analyses for a selection of the TACLeBench benchmarks.
    All bounds are normalized to the WCET bounds obtained with no persistence analysis at all, \ie, only running the traditional age-based must and may analysis~\cite{Ferdinand97}.
    For instance, a value of~$0.8$ means that the WCET bound is improved by $\SI{20}{\percent}$ using the respective persistence analysis compared with the WCET bound obtained without any persistence analysis.
    A general observation is that persistence analysis is often important to obtain precise WCET bounds with improvements of more than $\SI{20}{\percent}$ in several cases.

    We omitted a total of $27$~benchmarks from the figure, among which $5$ showed no differences between the analyses at all (as in \texttt{insertsort}); for $10$~omitted benchmarks, only $\cmust$ performed significantly worse than the rest (as in \texttt{lift}); and $12$~benchmarks only showed negligible differences (as in \texttt{pm}).
    An unabridged chart can be found in the \ifthenelse{\isundefined\istechreport}{technical report}{appendix of this technical report}.

    The chart shows that, in practice, the exact persistence analysis is only slightly more precise than existing inexact approaches.
    In most cases all persistence analyses perform almost identically (in particular in most of the omitted benchmarks), the only outlier being the \cmust analysis, \eg, in case of \texttt{lift} or \texttt{ndes}.
    Even the cheapest \gcs analysis usually performs similar to the exact analysis, except for \texttt{cjpeg\_wrbmp}.
    There are seven benchmarks in which the exact analysis obtains strictly better WCET bounds than all other analyses: the benchmarks \texttt{adpcm\_enc} ($\SI{0.2}{\percent}$), \texttt{epic} ($\SI{0.0006}{\percent}$), \texttt{h264\_dec} ($\SI{0.04}{\percent}$), \texttt{huff\_dec} ($\SI{0.6}{\percent}$), \texttt{ndes} ($\SI{0.1}{\percent}$), \texttt{pm} ($\SI{0.007}{\percent}$), and \texttt{powerwindow} ($\SI{2.4}{\percent}$).
    The numbers in parentheses indicate the marginal improvements in terms of the computed WCET bound.

    At its default settings, \LLVMTA{} heuristically performs loop peeling, \ie, it distinguishes the initial loop iteration from the following loop iterations using trace partitioning~\cite{Mauborgne2005}.
    The motivation for loop peeling is that the memory blocks used in the loop body miss the cache in the loop's first iteration, while the same memory blocks hit the cache in subsequent iterations.
    Without loop peeling, \ie, if the loop is considered as a whole, regular must and may analysis cannot classify such accesses.
    On the other hand, persistence analysis can cover such cases without loop peeling.
    Thus, to stress test the persistence analyses further, we performed another evaluation in which we deactivated loop peeling.
    As expected, without loop peeling, the average improvement of the exact persistence analysis over the plain must and may analysis in terms of WCET bounds rises to $\SI{52}{\percent}$---compared to $\SI{17}{\percent}$ with loop peeling.
    The exact analysis, however, is again on par with the previously known persistence analyses showing improvements in the range of only $\SI{0.2}{\percent}$ to $\SI{0.3}{\percent}$ in most cases.
    An interesting insight from this experiment is that the $\gcs$ analysis performs significantly worse relative to the remaining analyses once loop peeling is deactivated.
    This is likely due to the fact, that $\gcs$ is the only analysis whose analysis information for a block~$b$ is not ``reset'' upon an access to $b$ itself.
    Loop peeling conceals this limitation because the must analysis is able to classify many accesses as always hit for all but the first loop iteration.

    The execution time of a program can be seen as the sum of computation times and memory access latencies.
    The memory access latencies can further be split into contributions from data accesses and from instruction fetches.
    To focus the evaluation on the number of cache misses, we performed an analysis that separately bounds the maximum number of instruction and data cache misses.
    This experiment reveals that the observed differences between the persistence analyses are mainly due to the data cache.

\subsection{Analysis Cost}

    \begin{figure}[t]
        \centering%
        \begin{subfigure}[b]{\linewidth}
            \centering
            \scalebox{0.65}{\input{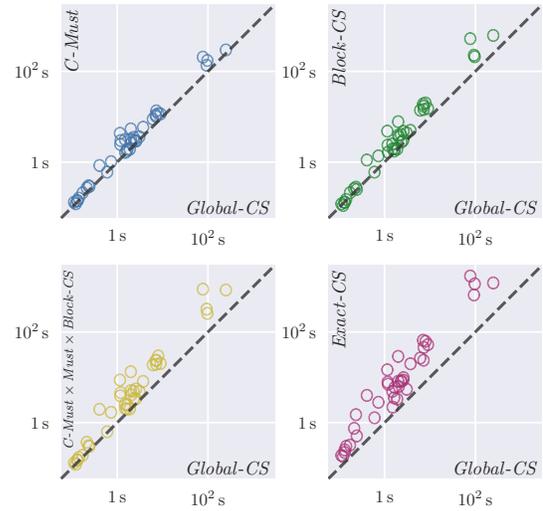}}
            \vspace{-2mm}
            \caption{Time}
            \label{fig:scatter-time}
            \vspace{2mm}
        \end{subfigure}
        \begin{subfigure}[b]{\linewidth}
            \hspace{4pt}
            \centering
            \scalebox{0.65}{\input{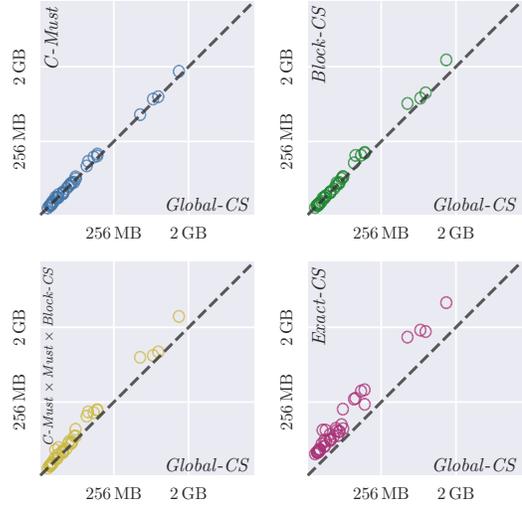}}
            \vspace{-2mm}
            \caption{Memory}
            \label{fig:scatter-mem}
        \end{subfigure}
        \caption{Run time and memory comparison of persistence analyses relative to $\gcs$.}
        \myvspace{-3mm}
    \end{figure}

    \begin{figure}
        \centering%
        \scalebox{0.65}{\input{scatter-8-32-cmmuele-exact.pgf}}
       \caption{Run time and memory comparison of $\ecs$ and $\cmmuele$.}
        \label{fig:scatter-cmmuele-exact}
    \end{figure}

    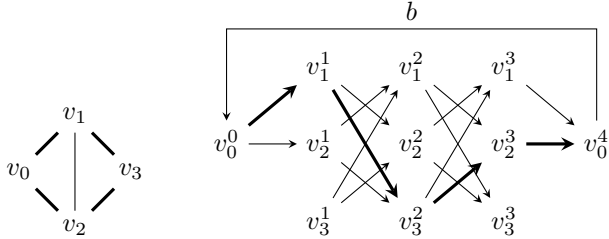
\begin{figure}[t]
        \centering
        \begin{subfigure}[t]{0.17\textwidth}
            \begin{center}
                \begin{tikzpicture}[node distance=3em]
                    \node (q0) { $v_0$ };
                    \node (q1) [above right of=q0] { $v_1$ };
                    \node (q3) [below right of=q1] { $v_3$ };
                    \node (q2) [below right of=q0]  { $v_2$ };
                    \path (q0) edge[very thick] (q1);
                    \path (q0) edge[very thick] (q2);
                    \path (q1) edge (q2);
                    \path (q1) edge[very thick] (q3);
                    \path (q2) edge[very thick] (q3);
                \end{tikzpicture}
            \end{center}
            \caption{Graph with Hamiltonian circuit (thick).}
            \label{subfig:ham}
        \end{subfigure}
        \begin{subfigure}[t]{0.31\textwidth}
            \begin{center}
                \begin{tikzpicture}[->, node distance=3.5em, auto]
                    \node (q0s)  { $v_0^0$ };

                    \node (q2_1) [right of=q0s]  { $v_2^1$ };
                    \node (q1_1) [above of=q2_1,yshift=-2mm] { $v_1^1$ };
                    \node (q3_1) [below of=q2_1,yshift=2mm] { $v_3^1$ };

                    \node (q1_2) [right of=q1_1] { $v_1^2$ };
                    \node (q2_2) [below of=q1_2,yshift=2mm]  { $v_2^2$ };
                    \node (q3_2) [below of=q2_2,yshift=2mm] { $v_3^2$ };

                    \node (q1_3) [right of=q1_2] { $v_1^3$ };
                    \node (q2_3) [below of=q1_3,yshift=2mm]  { $v_2^3$ };
                    \node (q3_3) [below of=q2_3,yshift=2mm] { $v_3^3$ };

                    \node (q0e) [right of=q2_3] { $v_0^4$ };

                    \node (aboveright) [above of=q0e, yshift=3mm] {};
                    \node (aboveleft) [above of=q0s, yshift=3mm] {};

                    \path[>=stealth] (q0s) edge[very thick] (q1_1);
                    \path[>=stealth] (q0s) edge (q2_1);

                    \path[>=stealth] (q1_1) edge (q2_2);
                    \path[>=stealth] (q1_1) edge[very thick] (q3_2);
                    \path[>=stealth] (q2_1) edge (q3_2);
                    \path[>=stealth] (q2_1) edge (q1_2);
                    \path[>=stealth] (q3_1) edge (q1_2);
                    \path[>=stealth] (q3_1) edge (q2_2);

                    \path[>=stealth] (q1_2) edge (q2_3);
                    \path[>=stealth] (q1_2) edge (q3_3);
                    \path[>=stealth] (q2_2) edge (q3_3);
                    \path[>=stealth] (q2_2) edge (q1_3);
                    \path[>=stealth] (q3_2) edge (q1_3);
                    \path[>=stealth] (q3_2) edge[very thick] (q2_3);

                    \path[>=stealth] (q1_3) edge (q0e);
                    \path[>=stealth] (q2_3) edge[very thick] (q0e);

                    \draw[>=stealth] (q0e) --  (aboveright.center) -- node[above] {$b$}  (aboveleft.center) --  (q0s);
                \end{tikzpicture}
            \end{center}
            \caption{%
                Control flow graph obtained by the reduction.
                Edge labels not shown, except for the back edge labeled~$b$, which is to be classified.
                The thick path corresponds to the Hamiltonian circuit of the graph in~(a).
            }
        \end{subfigure}

        \caption{Reduction from the Hamiltonian circuit problem.}
        \label{fig:exists-miss-reduction}
    \end{figure}

    This section evaluates the analysis cost of the different persistence analyses in terms of analysis run time and memory consumption.
    The same set of persistence analyses as in the previous section is evaluated.
    Additional experimental results may be found in the \ifthenelse{\isundefined\istechreport}{technical report}{appendix of this technical report}.

    The run time (Figure~\ref{fig:scatter-time}) and memory consumption (Figure~\ref{fig:scatter-mem}) results are visualized in scatter plots.
    In each of the scatter plots the horizontal axis corresponds to the value obtained for the simplest and presumably cheapest analysis $\gcs$.
    These values are compared with the remaining analyses on the vertical axis, respectively.
    A logarithmic scale is used for all scatter plots because the measured numbers vary greatly in size.
    Unsurprisingly, a general trend is that the exact analysis is the most expensive one regarding both analysis run time and memory consumption.
    However, even compared with the cheapest analysis \gcs, the memory overhead of the exact analysis is less than $3\times$ for all benchmarks and the analysis time is at most $23\times$ higher.

    In Figure~\ref{fig:scatter-cmmuele-exact}, the exact analysis is compared directly with the most precise analysis from the literature, $\cmmuele$.
    The data shows that the exact analysis is on the average $2\times$  slower and needs about $1.6\times$ more memory, which is indicated by the blue lines in the figure.

\section{Persistence Analysis is NP-complete}\label{sec:nphard}

    In this section, we show that persistence analysis is NP-complete.
    The \emph{persistence problem} is defined as follows: given a control-flow graph $G = \round{V, E, i}$, a designated memory block $b$, and a cache size $k$, is there a path through~$G$ that yields an access trace that results in more than one miss upon accesses to block~$b$ in a fully-associative LRU cache of size~$k$?

    \begin{theorem}\label{th:exists-miss-acyclic-np-complete}
        The \emph{persistence problem} is NP-complete.
    \end{theorem}

    \begin{proof}
        First, we show that the problem is indeed in NP.
        To this end, we show that if there is a witness path~$\pi$ that shows that block~$b$ is not persistent, then there is also a \emph{short} witness path $\pi'$, \ie, a witness path of length polynomial in the size of the control-flow graph~$G$.
        This proves that the problem is in NP, because a non-deterministic algorithm could first guess and then verify this witness path in polynomial time.

        Let $\pi$ be an arbitrary witness path through $G$ containing at least two accesses to block~$b$ that are misses in an LRU cache of size~$k$.
        Then $\pi$ can be decomposed as follows: $\pi = \textit{pre} \circ \round{s_i, b, s_{i+1}} \circ \textit{mid} \circ \round{s_j, b, s_{j+1}} \circ \textit{post}$, where the transition $\round{s_j, b, s_{j+1}}$ corresponds to the \emph{second miss} to $b$ among all accesses to $b$ in $\pi$, and \textit{mid} does not contain accesses to $b$, \ie, the transition $\round{s_i, b, s_{i+1}}$ corresponds to the last access to $b$ before the second miss to $b$ in $\pi$.

        Clearly, the suffix \textit{post} can be removed from $\pi$, and the resulting path is still a witness path.
        Next, we argue that \textit{mid} can be replaced by $\textit{mid}'$, such that $\abs{\textit{mid}'} < \abs{V}\cdot\abs{E}$, maintaining that the subsequent transition $\round{s_j, b, s_{j+1}}$ results in a miss:
        To this end, \textit{mid} is further decomposed into $\textit{mid} = \textit{mid}_1 \circ \textit{mid}_2 \circ \dots \circ \textit{mid}_n$, where each $\textit{mid}_i$ starts with an access to a memory block that was not accessed previously in $\textit{mid}$.
        Thus, the number of subpaths $n$ is the number of distinct memory blocks accessed on the path \textit{mid}.
        Clearly, $n < \abs{E}$.
        Each $\textit{mid}_i$ can be replaced by a $\textit{mid}_i'$, such that $\abs{\textit{mid}_i'} \leq \abs{V}$:
        Such a $\textit{mid}_i'$ can be obtained by keeping the first transition of $\textit{mid}_i$ and then removing from $\textit{mid}_i$ any cycles, \ie, subpaths starting and ending in the same node.
        By construction, $\textit{mid}' = \textit{mid}_1' \circ \textit{mid}_2' \circ \dots \circ \textit{mid}_n'$ does not contain accesses to $b$ and consists of accesses to at least as many distinct memory blocks as $\textit{mid}$.
        Finally, \textit{pre} can be replaced by the shortest path~$\textit{pre}'$ in $G$ from the initial location~$i$ to $s_i$.
        Clearly, $\abs{\textit{pre}'} \leq \abs{V}$.
        Also, the first access to $b$ in $\textit{pre}' \circ \round{s_i, b, s_{i+1}}$, which must exist due to the final transition~$\round{s_i, b, s_{i+1}}$, results in a miss.

        Thus, the path $\pi' = \textit{pre}' \circ \round{s_i, b, s_{i+1}} \circ \textit{mid}' \circ \round{s_j, b, s_{j+1}}$ is also a witness to the fact that $b$ is not persistent, and its length is bounded by $\abs{V}+\abs{V}\cdot\abs{E} + 2$, \ie, it is polynomial in the size of the control-flow graph $G$.

        Now we show that the persistence problem is NP-hard. 
        This part of the proof is analogous to Touzeau \etal's proof that LRU must analysis is NP-hard~\cite{Touzeau2019}.

        We reduce the Hamiltonian circuit problem to the persistence problem.
        Let $\round{V, E}$ be a graph, let $n = \abs{V}$, $V = \set{v_0,\dots,v_{n-1}}$.
        We construct a control-flow graph $G = \round{V', E', i}$ for cache persistence analysis as follows:
        \begin{itemize}
            \item
                Create two copies $v_0^0$ and $v_0^n$ of $v_0$ in $V'$.
            \item
                For each $v_i$, $i \geq 1$, create $\abs{V}-1=n-1$ copies $v_i^j$, $1 \leq j < n$ in $V'$.
                This arranges these vertices in layers indexed by~$j$.
            \item
                For each pair $v_j^l$, $v_{j'}^{l+1}$ of nodes in consecutive layers, create an edge in $E'$, labeled by the address~$j'$, if and only if there is an edge $\round{j,j'}$ in~$E$.
            \item
                The initial control location $i$ is $v_0^0$.
        \end{itemize}
        See Figure~\ref{fig:exists-miss-reduction} for an example.
        There is a Hamiltonian circuit in $\round{V,E}$ if and only if there is a path in $G$ from $v_0^0$ to $v_0^n$ such that no edge label is repeated,
        thus if and only if there exists a path from $v_0^0$ to $v_0^n$ with at least $n$ distinct edge labels.

        Now assume an edge going from $v_0^n$ back to $v_0^0$ labeled with the fresh memory block~$b$.
        This memory block $b$ is the one to classify.
        For cache size $n$ there exists a path resulting in two or more misses to~$b$ if and only if there is a path from $v_0^0$ to $v_0^n$ with at least $n$ distinct edge labels, corresponding to a Hamiltonian circuit in the graph~$\round{V, E}$.
    \end{proof}
   
   It is important to point out that the above NP-hardness proof critically relies on the cache size being an input parameter.
   In fact, it turns out to be possible to devise a persistence analysis that is exponential in the cache size but polynomial in the size of the control-flow graph~\cite{Dell2019} based on recent results in theoretical computer science~\cite{Brand2018}.

\section{Conclusions and Future Work}\label{sec:conclusions}

    We have shown that it is possible to perform exact cache persistence analysis for caches with LRU replacement at a reasonable analysis cost.
    To this end, we introduced a sequence of exact abstractions, exploiting monotonicity properties inherent to LRU replacement; followed by an efficient implementation based on zero-suppressed binary decision diagrams (ZDDs).
    In addition, we introduced novel techniques to efficiently deal with uncertainty arising in the context of data cache analysis.

    The motto of our paper could be: ``in practice, theory and practice are different'', as the following findings demonstrate:
    Our experimental evaluation reveals that the new exact analysis is only moderately more costly than existing inexact analyses; in particular, in our experiments it does not exhibit exponential complexity in terms of the input size.
    This is in spite of the fact that its worst-case complexity is indeed exponential and that persistence analysis is NP-hard, as we show in Section~\ref{sec:nphard}.
    Similarly, while even the most precise existing persistence analyses are not exact in theory, our experiments show that they are close to exact in practice.

\section*{Acknowledgements}
This work was supported by the Deutsche Forschungsgemeinschaft as part of the project PEP -- 289264719.
We thank the anonymous reviewers for their helpful comments.

\begin{techreport}
    \clearpage
    \section{Appendix}

    \begin{theorem}[Exactness of Persistence Analysis]\label{thm:exact-relative}
        Let $\mathcal{P}$ be a persistence analysis over a finite abstract domain~$\abstraces[\mathcal{P}]{}$ that satisfies the equations of Theorem~\ref{thm:exact}.
        Furthermore let $\mathcal{Q}$ be a persistence analysis over a finite abstract domain~$\abstraces[\mathcal{Q}]{}$ defined relative to $\mathcal{P}$ by an abstraction function $\abstraction[\mathcal{Q}]{}$.
        The persistence analysis $\mathcal{Q}$ is \emph{exact} if:
        \begin{align*}
            \forall T, b &: \abstraction[\mathcal{Q}]{}\round[\big]{\update[\mathcal{P}]{T, b}} = \update[\mathcal{Q}]{\abstraction[\mathcal{Q}]{T}, b} \\
            \forall I, T_i &: \abstraction[\mathcal{Q}]{}\round[\Big]{\sideset{}{_\mathcal{P}}\bigsqcup_{i \in I} T_i} = \sideset{}{_\mathcal{Q}}\bigsqcup_{i \in I} \abstraction[\mathcal{Q}]{T_i}
        \end{align*}
        and the abstraction preserves the persistence classification:
        \begin{align*}
            \forall T, b : \classify[\mathcal{P}]{T, b} \leftrightarrow \classify[\mathcal{Q}]{\abstraction[\mathcal{Q}]{T}, b}
        \end{align*}
    \end{theorem}
    \begin{proof}
        Composing the abstraction functions~$\abstraction[\mathcal{P}]{}$ and ~$\abstraction[\mathcal{Q}]{}$ we obtain the abstraction function $\abstraction[\mathcal{Q}\circ\mathcal{P}]{} \coloneqq \abstraction[\mathcal{Q}]{} \circ \abstraction[\mathcal{P}]{}$ that relates the persistence analysis~$\mathcal{Q}$ directly with the concrete trace semantics.
        If we show that the equations of Theorem~\ref{thm:exact} are satisfied for $\abstraction[\mathcal{Q}\circ\mathcal{P}]{}$, then the exactness of $\mathcal{Q}$ follows.
        \begin{enumerate}
            \setlength{\itemsep}{6pt}
            \item
                $\forall T, b : \abstraction[\mathcal{Q}\circ\mathcal{P}]{\mi{update}\round{T, b}} = \update[\mathcal{Q}]{\abstraction[\mathcal{Q}\circ\mathcal{P}]{T}, b}$
            \item
                $\forall I, T_i : \abstraction[\mathcal{Q}\circ\mathcal{P}]{}\round[\big]{\bigcup\limits_{i \in I} T_i} = \sideset{}{_\mathcal{Q}}{\textstyle\bigsqcup}\limits_{i \in I} \abstraction[\mathcal{Q}\circ\mathcal{P}]{T_i}$
            \item
                $\forall T, b : \mi{persistent}\round{T, b} \leftrightarrow \classify[\mathcal{Q}]{\abstraction[\mathcal{Q}\circ\mathcal{P}]{T}, b}$
        \end{enumerate}
        The proof is straightforward and follows immediately from the equations of Theorem~\ref{thm:exact} for $\mathcal{P}$ and the premises of this theorem.
    \end{proof}

    \thmexactecsb*
    \begin{proof}
        It suffices to prove the equations of Theorem~\ref{thm:exact}.
        \begin{enumerate}
            \setlength{\itemsep}{6pt}
            \item
                $\forall T, b : \abstraction[\xxecsb]{\mi{update}\round{T, b}} = \update[\xxecsb]{\abstraction[\xxecsb]{T}, b}$ \\[1ex]
                Let $T$ be a set of traces and $b\in\blocks$.
                By unfolding definitions, the term $\abstraction[\xxecsb]{\mi{update}\round{T, b}}$ reduces to:
                \begin{align*}
                    \lam{b'}{\bigcup_{\tau\in T} \begin{cases}
                        \set{\set{b}}                                                               & \text{if $b' = b$} \\
                        \set{s\cup \set{b} \where s\in \abstraction[\xxecsb]{\set{\tau}}\round{b'}} & \text{otherwise} \\
                    \end{cases}}
                \end{align*}
                Next, the union over all traces can be drawn in:
                \begin{align*}
                    \lam{b'}{\begin{cases}
                        \set{\set{b}}                                                                                   & \text{if $b' = b$} \\
                        \set[\big]{s\cup \set{b} \where s\in \bigcup_{\tau\in T} \abstraction[\xxecsb]{\set{\tau}}\round{b'}} & \text{otherwise} \\
                    \end{cases}}
                \end{align*}
                This is equivalent to $\update[\xxecsb]{\abstraction[\xxecsb]{T}, b}$ by definition.
            \item
                $\forall T_i : \abstraction[\xxecsb]{}\round[\big]{\bigcup\limits_{i \in\NN_0} T_i} = \sideset{}{_\xxecsb}{\textstyle\bigsqcup}\limits_{i \in\NN_0} \abstraction[\xxecsb]{T_i}$ \\[1ex]
                The claim follows trivially from the definition of $\abstraction[\xxecsb]{}$.
            \item
                $\forall T, b : \mi{persistent}\round{T, b} \leftrightarrow \classify[\xxecsb]{\abstraction[\xxecsb]{T}, b}$ \\[1ex]
                Let $T$ be an arbitrary set of traces and $b\in\blocks$.
                The left hand side is equivalent to $\forall \tau\in T : \mi{persistent}\round{\tau, b}$ by definition while the right hand side can be reformulated as $\forall \tau\in T : \max\set{\abs{s} \where s\in \abstraction[\xxecsb]{\tau}\round{b}} \leq k$.
                It is therefore sufficient to apply Lemma~\ref{lem:classify} on all $\tau \in T$. \qedhere
        \end{enumerate}
    \end{proof}

    \begin{lemma}\label{lem:abstraction-empty}
        \begin{align*}
            \forall \tau, b : \round{\forall 0\leq i < \abs{\tau} : \tau_i \neq b} \rightarrow \abstraction[\xxecsb]{\tau}\round{b} = \emptyset
        \end{align*}
        \begin{proof}
            Let $\tau$ be a trace and $b\in\blocks$.
            We are always in the second case of $\update[\xxecsb]{}$ and propagate the initial information $\abstraction[\xxecsb]{\epsilon}\round{b} = \emptyset$.
        \end{proof}
    \end{lemma}

    \begin{lemma}\label{lem:abstraction-cs}
        \begin{align*}
            \forall \tau, b : \round{\exists 0\leq i < \abs{\tau} : \tau_i = b} \rightarrow \abstraction[\xxecsb]{\tau}\round{b} = \set{\mi{CS}\round{\tau, b}}
        \end{align*}
        \begin{proof}
            Let $\tau$ be a trace, $b\in\blocks$ and $m \coloneqq \max\set{0\leq i <\abs{\tau} \where \tau_i = b}$.
            We know that $\abstraction[\xxecsb]{\tau_0 \circ\cdots\circ \tau_m}\round{b} = \set{\set{b}}$ by the definition of $\update[\xxecsb]{}$.
            Furthermore, $\set{\mi{CS}\round{\tau, b}} = \set{t_j \where j \geq m}$.
            Now, all upcoming updates just add some block in $\mi{CS}\round{\tau, b}$ to $\set{b}$ and the claim follows.
        \end{proof}
    \end{lemma}

    \begin{lemma}\label{lem:classify}
        \begin{align*}
            \forall \tau, b : \mi{persistent}\round{\tau, b} \leftrightarrow \classify[\xxecsb]{\abstraction[\xxecsb]{\set{\tau}}, b}
        \end{align*}
        \begin{proof}
            Let $\tau$ be a trace and $b\in\blocks$.
            The left hand side reduces to ${\round{\exists 0\leq i < \abs{\tau} : \tau_i = b} \rightarrow \abs{\mi{CS}\round{\tau, b}} \leq k}$ and the right hand side to $\max\set{\abs{s} \where s\in \abstraction[\xxecsb]{\set{\tau}}\round{b}} \leq k$ by unfolding all definitions.
            Applying Lemmas~\ref{lem:abstraction-empty} and \ref{lem:abstraction-cs} completes the proof.
        \end{proof}
    \end{lemma}

    \thmexactecsm*
    \begin{proof}
        It suffices to prove the equations of Theorem~\ref{thm:exact-relative}.
        \begin{enumerate}
            \setlength{\itemsep}{6pt}
            \item
                $\forall T, b : \abstraction[\xxecsm]{}\round[\big]{\update[\xxecsb]{T, b}} = \update[\xxecsm]{\abstraction[\xxecsm]{T}, b}$ \\[1ex]
                Let $T\in \abstraction[\xxecsb]{}$ be an abstract trace and $b\in\blocks$.
                We show that both functions agree on all $b'\in\blocks$.
                The case $b' = b$ is trivial.\\
                If $b' \neq b$, the left hand side $\abstraction[\xxecsm]{}\round[\big]{\update[\xxecsb]{T, b}}\round{b'}$ reduces to:
                \begin{align*}
                    \mi{maxSet}\round{\set{s \cup \set{b} \where s\in T\round{b'}}}
                \end{align*}
                while the right hand side $\update[\xxecsm]{}\round[\big]{\abstraction[\xxecsm]{T}, b}\round{b'}$ reduces to:
                \begin{align*}
                    \mi{maxSet}\round{\set{s \cup \set{b} \where s\in \mi{maxSet}\round{T\round{b'}}}}
                \end{align*}
                The missing step is closed by Lemma~\ref{lem:maxSet-dotproduct}.
            \item
                $\forall I, T_i : \abstraction[\xxecsm]{}\round[\big]{\sideset{}{_\xxecsb}{\textstyle\bigsqcup}\limits_{i\in I} T_i} = \sideset{}{_\xxecsm}{\textstyle\bigsqcup}\limits_{i\in I} \abstraction[\xxecsm]{T_i}$ \\[1ex]
                Let $I\subseteq\NN_0$ and $T_i\in \abstraction[\xxecsb]{}$ be arbitrary abstract traces.
                We show that both functions agree on all $b\in\blocks$.
                By definition, we can transform $\abstraction[\xxecsm]{}\round[\big]{\sideset{}{_\xxecsb}{\textstyle\bigsqcup}\limits_{i\in I} T_i}\round{b}$ to $\mi{maxSet}\round[\big]{\bigcup\limits_{i\in I} T_i\round{b}}$.
                This is equal to $\mi{maxSet}\round[\big]{\bigcup\limits_{i\in I} \mi{maxSet}\round{T_i\round{b}}}$ by Lemma~\ref{lem:maxSet-union} which is equal to $\round[\big]{\sideset{}{_\xxecsm}{\textstyle\bigsqcup}\limits_{i\in I} \abstraction[\xxecsm]{T_i}}\round{b}$ by definition.
            \item
                $\forall T, b : \classify[\xxecsb]{T, b} \leftrightarrow \classify[\xxecsm]{\abstraction[\xxecsm]{T}, b}$ \\[1ex]
                Let $T\in \abstraction[\xxecsb]{}$ be an abstract trace and $b\in\blocks$.
                The claim is proven with Lemma~\ref{lem:maxSet-cardinality} after unfolding all definitions. \qedhere
        \end{enumerate}
    \end{proof}

    \begin{lemma}\label{lem:maxSet-dotproduct}
        \begin{multline*}
            \forall A, b : \mi{maxSet}\round{\set{s\cup \set{b} \where s\in A}} =\\
            \mi{maxSet}\round{\set{s\cup \set{b} \where s\in \mi{maxSet}(A)}}
        \end{multline*}
        \begin{proof}
            Let $A$ be a set and $b\in\blocks$.
            We prove the claim by showing mutual inclusion.
            \begin{itemize}
                \item[``$\subseteq$'':]
                    Let $t\in \mi{maxSet}\round{\set{s\cup \set{b} \where s\in A}}$, \ie, $t = s\cup \set{b}$ for some $s\in A$.
                    We have to prove two statements:
                    \begin{enumerate}
                        \item
                            $t\in \set{s\cup \set{b} \where s\in \mi{maxSet}\round{A}}$ \\
                            In the case $t\in A$, we claim that $t\in\mi{maxSet}\round{A}$.
                            By assuming the contrary, we immediately get a contradiction with $t\in \mi{maxSet}\round{\set{s\cup \set{b} \where s\in A}}$. \\
                            If $t\notin A$, \ie, $t = s\uplus \set{b}$, we claim that $s\in\mi{maxSet}\round{A}$.
                            Assume the contrary, \ie, there is $s'\in A$, $s\subsetneq s'$.
                            Note that $s' \neq t$ as $t\notin A$.
                            Then, $s'\cup\set{b} \supsetneq t$ contradicts $t\in \mi{maxSet}\round{\set{s\cup \set{b} \where s\in A}}$.
                        \item
                            $\neg\exists s' \in \set{s\cup \set{b} \where s\in \mi{maxSet}\round{A}} : t\subsetneq s'$ \\
                            If we assume otherwise, we get a contradiction to $t\in \mi{maxSet}\round{\set{s\cup \set{b} \where s\in A}}$.
                    \end{enumerate}
                \item[``$\supseteq$'':]
                    Trivial, as $\mi{maxSet}\round{A} \subseteq A$. \qedhere
            \end{itemize}
        \end{proof}
    \end{lemma}

    \begin{lemma}\label{lem:maxSet-union}
        \begin{multline*}
            \forall I, A_i : \mi{maxSet}\round[\Big]{\bigcup_{i\in I} A_i} = \mi{maxSet}\round[\Big]{\bigcup_{i\in I} \mi{maxSet}(A_i)}
        \end{multline*}
        \begin{proof}
            Let $I\subseteq \NN_0$, $A_i$ be arbitrary sets and $\widehat{A}_i\coloneqq \mi{maxSet}\round{A_i}$.
            We prove the claim by showing mutual inclusion.
            \begin{itemize}
                \setlength{\itemsep}{6pt}
                \item[``$\subseteq$'':]
                    Let $s\in\mi{maxSet}\round*{\bigcup_{i\in I} A_i}$, \ie, $s\in \bigcup_{i\in I} A_i$ and $\neg \exists s'\in \bigcup_{i\in I} A_i : s\subsetneq s'$.
                    It is easy to see that $s\in \widehat{A}_j$ for some $j\in I$, and hence, $s\in \bigcup_{i\in I} \widehat{A}_i$.
                    As $\round[\big]{\bigcup_{i\in I} \widehat{A}_i} \subseteq \round*{\bigcup_{i\in I} A_i}$, we get that $s\in \mi{maxSet}\round[\big]{\bigcup_{i\in I} \widehat{A}_i}$.
                \item[``$\supseteq$'':]
                    Trivial, as $\round[\big]{\bigcup_{i\in I} \widehat{A}_i} \subseteq \round*{\bigcup_{i\in I} A_i}$. \qedhere
            \end{itemize}
        \end{proof}
    \end{lemma}

    \begin{lemma}\label{lem:maxSet-cardinality}
        \begin{align*}
            \forall A : \max\set{\abs{x} \where x\in A} = \max\set{\abs{x} \where x\in \mi{maxSet}\round{A}}
        \end{align*}
        \begin{proof}
            Let $A$ be a set.
            We prove the equality by showing mutual less or equal relations.
            \begin{enumerate}
                \item[``$\leq$'':]
                    Let $n \coloneqq \max\set{\abs{x} \where x\in A}$ and $m\coloneqq \max\set{\abs{x} \where x\in \mi{maxSet}\round{A}}$.
                    Assume that $n > m$ and there is $a\in A$ with $\abs{a} = n$.
                    There cannot exist $b\in A$ with $a \subsetneq b$ because otherwise $\abs{b} > n$ is a contradiction.
                    But then, $a\in\mi{maxSet}\round{A}$ which contradicts $\abs{a} = n > m$.
                \item[``$\geq$'':]
                    Trivial, as $\mi{maxSet}\round{A} \subseteq A$. \qedhere
            \end{enumerate}
        \end{proof}
    \end{lemma}

    \thmexactecsc*
    \begin{proof}
        With the proof of Theorem~\ref{thm:exact-ecsm}, we have already shown that the domain of $\abstraction[\xxecsm]{}$ satisfies the equations of Theorem~\ref{thm:exact}.
        As $\abstraction[\xxecsc]{}$ is defined relative to this domain, is is sufficient to prove the equations of Theorem~\ref{thm:exact-relative}.
        \begin{enumerate}
            \setlength{\itemsep}{6pt}
            \item
                $\forall T, b : \abstraction[\xxecsc]{}\round[\big]{\update[\xxecsm]{T, b}} = \update[\xxecsc]{\abstraction[\xxecsc]{T}, b}$ \\[1ex]
                Let $T\in \abstraction[\xxecsm]{}$ be an abstract trace and $b\in\blocks$.
                We show that both functions agree on all $b'\in\blocks$.
                The case $b' = b$ is trivial.\\
                If $b' \neq b$, the left hand side $\abstraction[\xxecsc]{\update[\xxecsm]{T, b}}\round{b'}$ reduces to:
                \begin{align}
                    \mi{limit}\round{\mi{maxSet}\round{\set{s \cup \set{b} \where s\in T\round{b'}}}} \label{eq:proof-ecsc-1}
                \end{align}
                We have to show equality with:
                \begin{align}
                    \mi{limit}\round{\mi{maxSet}\round{\set{s \cup \set{b} \where s\in \mi{limit}\round{T\round{b'}}}}} \label{eq:proof-ecsc-2}
                \end{align}
                which is the right hand side $\update[\xxecsc]{\abstraction[\xxecsc]{T}, b}\round{b'}$ where all definitions are unfolded. \\
                We distinguish the two cases of $\mi{limit}\round{T\round{b'}}$.
                The case $\mi{limit}\round{T\round{b'}} = T\round{b'}$ is trivial.
                Therefore, assume $\exists s\in T\round{b'} : \abs{s} > k$.
                Equation~\ref{eq:proof-ecsc-2} reduces to $\set{\blocks}$ by definition as $\mi{limit}\round{T\round{b'}} = \set{\blocks}$.
                For Equation~\ref{eq:proof-ecsc-1}, the union with $\set{b}$ cannot decrease the cardinality and we get $\exists s'\in \set{s \cup \set{b} \where s\in T\round{b'}} : \abs{s'} > k$.
                The proof is completed by applying Lemma~\ref{lem:maxSet-cardinality} and the definition of $\mi{limit}$ which prove that~(\ref{eq:proof-ecsc-1}) is equal to $\set{\blocks}$, too.
            \item
                $\forall I, T_i : \abstraction[\xxecsc]{}\round[\big]{\sideset{}{_\xxecsm}{\textstyle\bigsqcup}\limits_{i\in I} T_i} = \sideset{}{_{\xxecsc}}{\textstyle\bigsqcup}\limits_{i\in I} \abstraction[\xxecsc]{T_i}$ \\[1ex]
                Let $I\subseteq\NN_0$ and $T_i\in \abstraction[\xxecsm]{}$ be arbitrary abstract traces.
                We show that both functions agree on all $b\in\blocks$.
                By definition, we can transform $\abstraction[\xxecsc]{}\round[\big]{\sideset{}{_\xxecsm}{\textstyle\bigsqcup}\limits_{i\in I} T_i}\round{b}$ to $\mi{limit}\round[\big]{\mi{maxSet}\round[\big]{\bigcup\limits_{i\in I} T_i\round{b}}}$.
                This is equal to $\mi{maxSet}\round[\big]{\bigcup\limits_{i\in I} \mi{limit}\round{T_i\round{b}}}$ by Lemma~\ref{lem:limit-union} which is equal to $\round[\big]{\sideset{}{_{\xxecsc}}{\textstyle\bigsqcup}\limits_{i\in I} \abstraction[\xxecsc]{T_i}}\round{b}$ by definition.
            \item
                $\forall T, b : \classify[\xxecsm]{T, b} \leftrightarrow \classify[\xxecsc]{\abstraction[\xxecsc]{T}, b}$ \\[1ex]
                Let $T\in \abstraction[\xxecsm]{}$ be an abstract trace and $b\in\blocks$.
                The claim is proven with Lemma~\ref{lem:limit-cardinality} after unfolding all definitions. \qedhere
        \end{enumerate}
    \end{proof}

    \begin{lemma}\label{lem:limit-union}
        \begin{multline*}
            \forall I, A_i : \mi{limit}\round[\Big]{\mi{maxSet}\round[\Big]{\bigcup_{i\in I} A_i}} =\\
            \mi{maxSet}\round[\Big]{\bigcup_{i\in I} \mi{limit}(A_i)}
        \end{multline*}
        \begin{proof}
            Let $I\subseteq\NN_0$ and $A_i$ be arbitrary sets.
            We prove the claim by showing mutual inclusion.
            \begin{itemize}
                \item[``$\subseteq$'':]
                    Let $s \in \mi{limit}\round[\big]{\mi{maxSet}\round[\big]{\bigcup_{i\in I} A_i}}$.
                    \begin{itemize}
                        \item
                            Assume $\exists s'\in \mi{maxSet}\round[\big]{\bigcup_{i\in I} A_i} : \abs{s'} > k$, \ie, $s=\blocks$.
                            W.\,l.\,o.\,g. let $s'\in A_j$ for some $j\in I$.
                            Then, $\mi{limit}\round{A_j} = \set{\blocks}$ and $s\in \set{\blocks} = \mi{maxSet}\round[\big]{\bigcup_{i\in I} \mi{limit}\round{A_i}}$.
                        \item
                            Otherwise, $s \in \mi{maxSet}\round[\big]{\bigcup_{i\in I} A_i}$ and $\neg\exists s'\in \mi{maxSet}\round[\big]{\bigcup_{i\in I} A_i} : \abs{s'} > k$.
                            Therefore, $\forall i\in I : \neg\exists s'\in A_i : \abs{s'} > k$, \ie, $\mi{limit}\round{A_i} = A_i$.
                    \end{itemize}
                \item[``$\supseteq$'':]
                    Let $s \in \mi{maxSet}\round[\big]{\bigcup_{i\in I} \mi{limit}\round{A_i}}$.
                    If $\forall i\in I: \neg\exists s'\in A_i : \abs{s'} > k$, then $s\in\mi{maxSet}\round[\big]{\bigcup_{i\in I} A_i}$.
                    Otherwise let w.\,l.\,o.\,g. $\mi{limit}\round{A_j} = \set{\blocks}$ for some $j\in I$.
                    Then, $s = \blocks\in \mi{limit}\round[\big]{\mi{maxSet}\round[\big]{\bigcup_{i\in I} A_i}}$.\qedhere
            \end{itemize}
        \end{proof}
    \end{lemma}

    \begin{lemma}\label{lem:limit-cardinality}
        \begin{multline*}
            \forall A, k : \max\set{\abs{x} \where x\in A}\leq k \leftrightarrow\\
            \max\set{\abs{x} \where x\in \mi{limit}\round{A}}\leq k
        \end{multline*}
        \begin{proof}
            Let $A$ be a set and $k\in\NN$.
            We prove the claim by showing mutual implication.
            \begin{enumerate}
                \item[``$\rightarrow$'':]
                    Let $\max\set{\abs{x} \where x\in A}\leq k$, \ie, $\neg\exists y\in A : \abs{y} > k$.
                    But then we have $\mi{limit}\round{A} = A$.
                \item[``$\leftarrow$'':]
                    Let $\max\set{\abs{x} \where x\in \mi{limit}\round{A}}\leq k$.
                    Assume $\mi{limit}\round{A} = \set{\blocks}$, then $\exists y\in A : \abs{y} > k$, a contradiction. \qedhere
            \end{enumerate}
        \end{proof}
    \end{lemma}

    \begin{figure*}[p]
        \begin{subfigure}[c]{0.5\textwidth}
            \centering%
            \scalebox{0.8}{\includegraphics{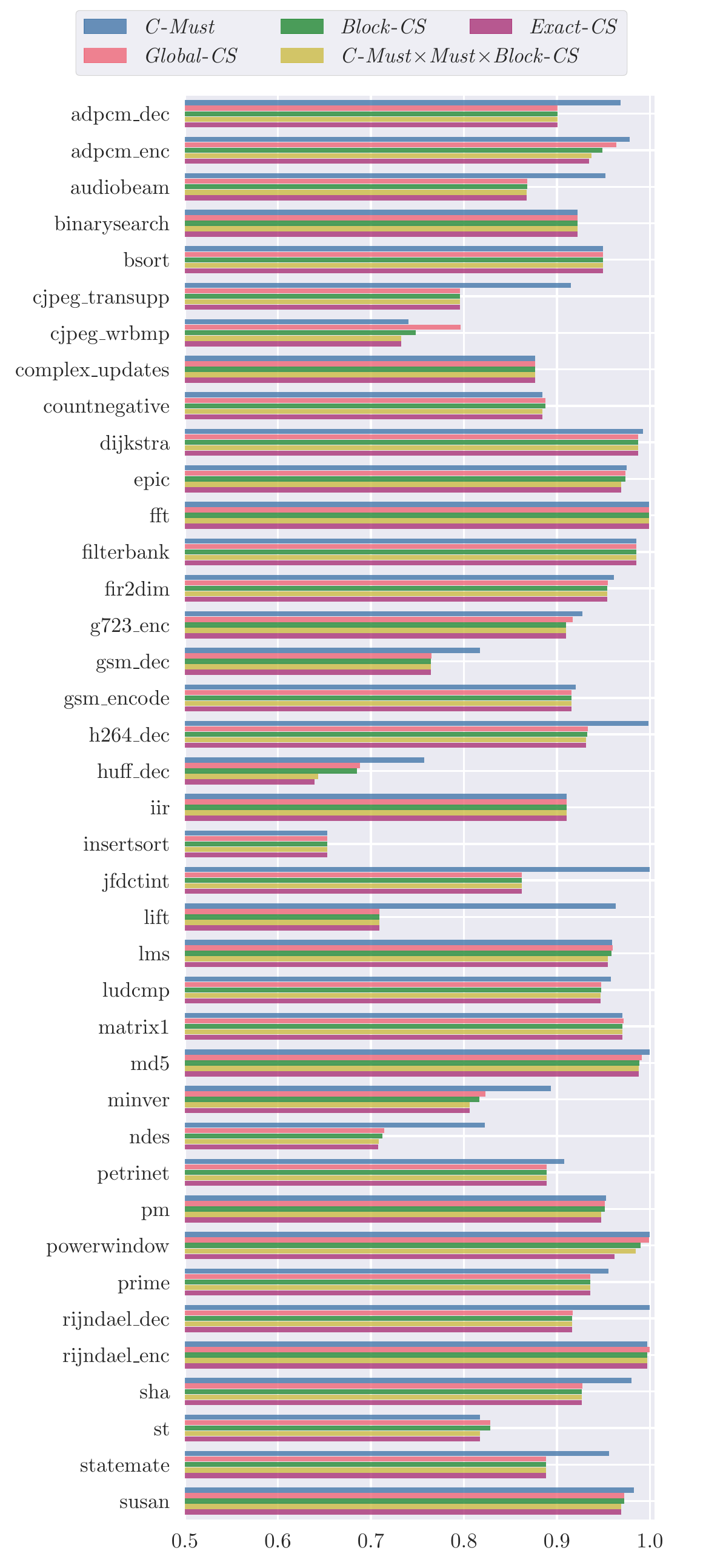}}
            \subcaption{\emph{Without} compiler optimizations.}
        \end{subfigure}
        \begin{subfigure}[c]{0.5\textwidth}
            \centering%
            \scalebox{0.8}{\includegraphics{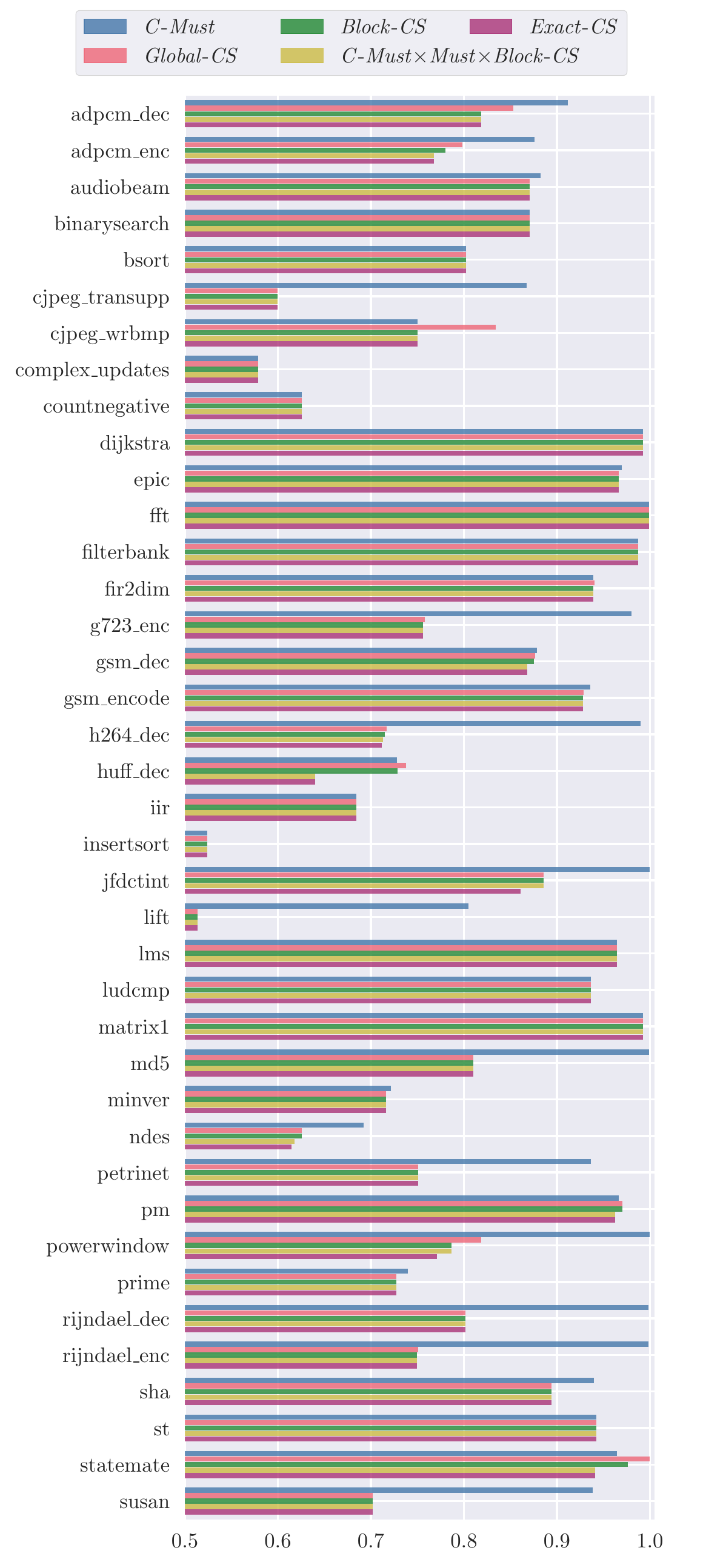}}
            \subcaption{\emph{With} compiler optimizations.}
        \end{subfigure}
        \caption{WCET ratios of the persistence analyses compared with performing no persistence analysis at all.\\Cache configuration: $32$ cache sets, $8$ ways, $16\,\mathrm{B}$ line size.}
    \end{figure*}

    \begin{figure*}[p]
        \begin{subfigure}[c]{0.5\textwidth}
            \centering%
            \scalebox{0.8}{\includegraphics{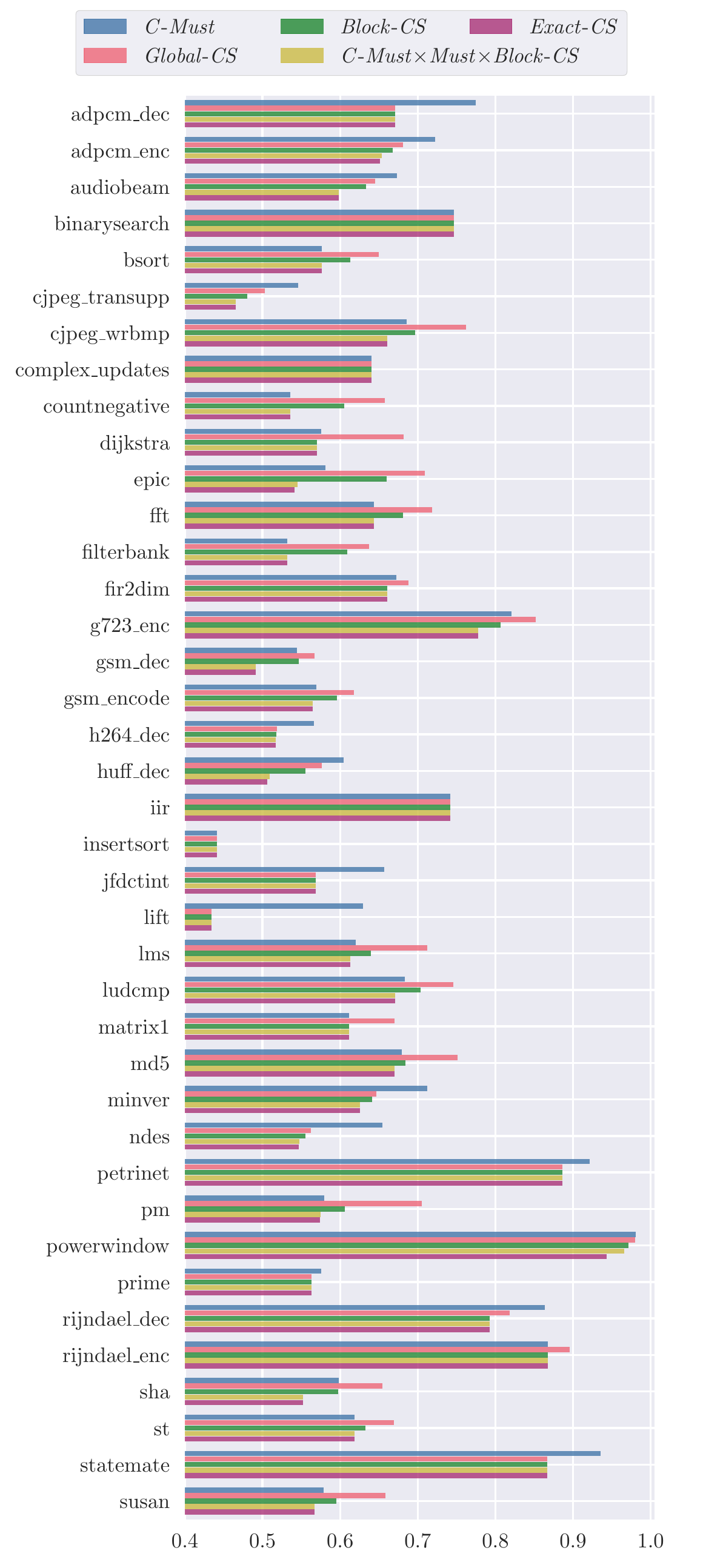}}
            \subcaption{\emph{Without} compiler optimizations.}
        \end{subfigure}
        \begin{subfigure}[c]{0.5\textwidth}
            \centering%
            \scalebox{0.8}{\includegraphics{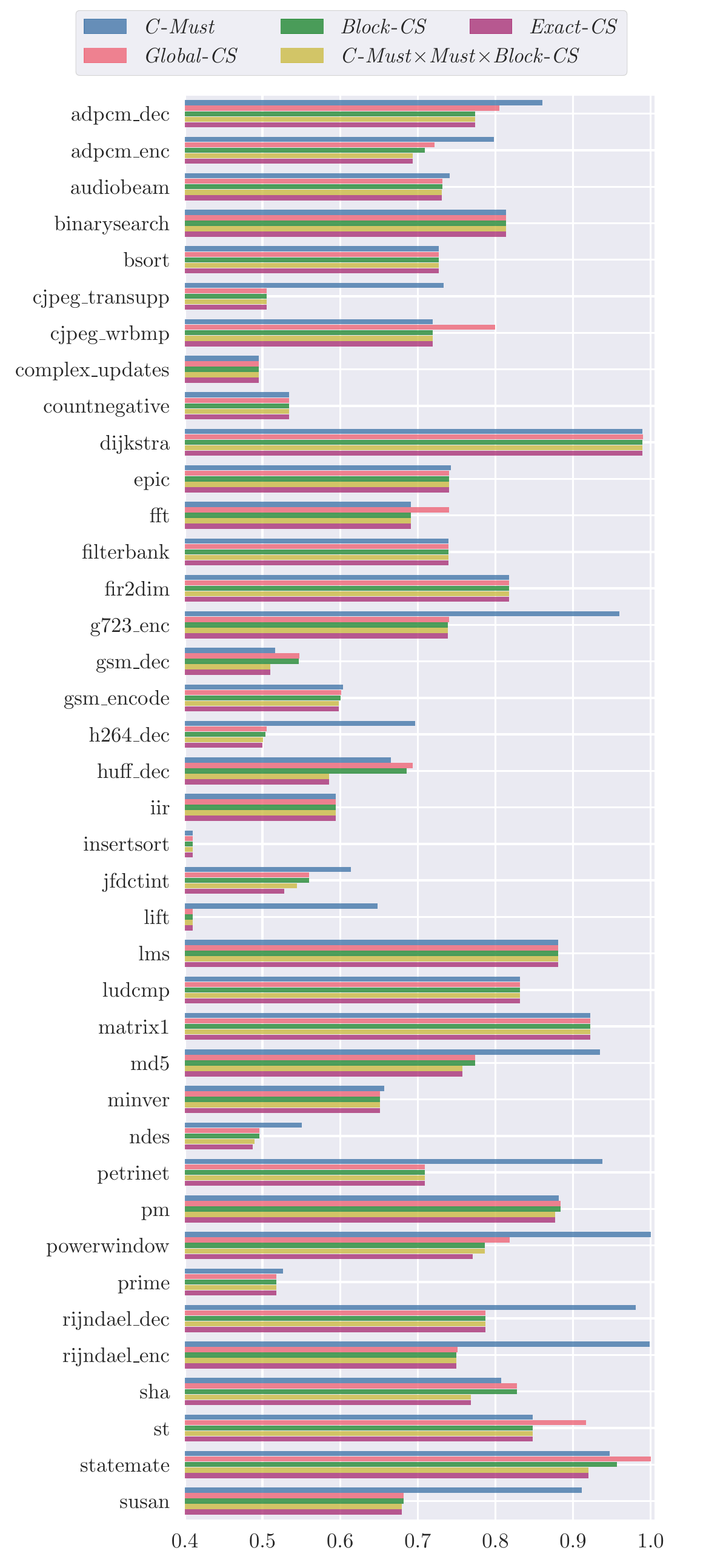}}
            \subcaption{\emph{With} compiler optimizations.}
        \end{subfigure}
        \caption{WCET ratios of the persistence analyses compared with performing no persistence analysis at all.\\Cache configuration: $32$ cache sets, $8$ ways, $16\,\mathrm{B}$ line size, \emph{loop peeling disabled}.}
    \end{figure*}

    \begin{figure*}[p]
        \begin{subfigure}[c]{0.5\textwidth}
            \centering%
            \scalebox{0.8}{\includegraphics{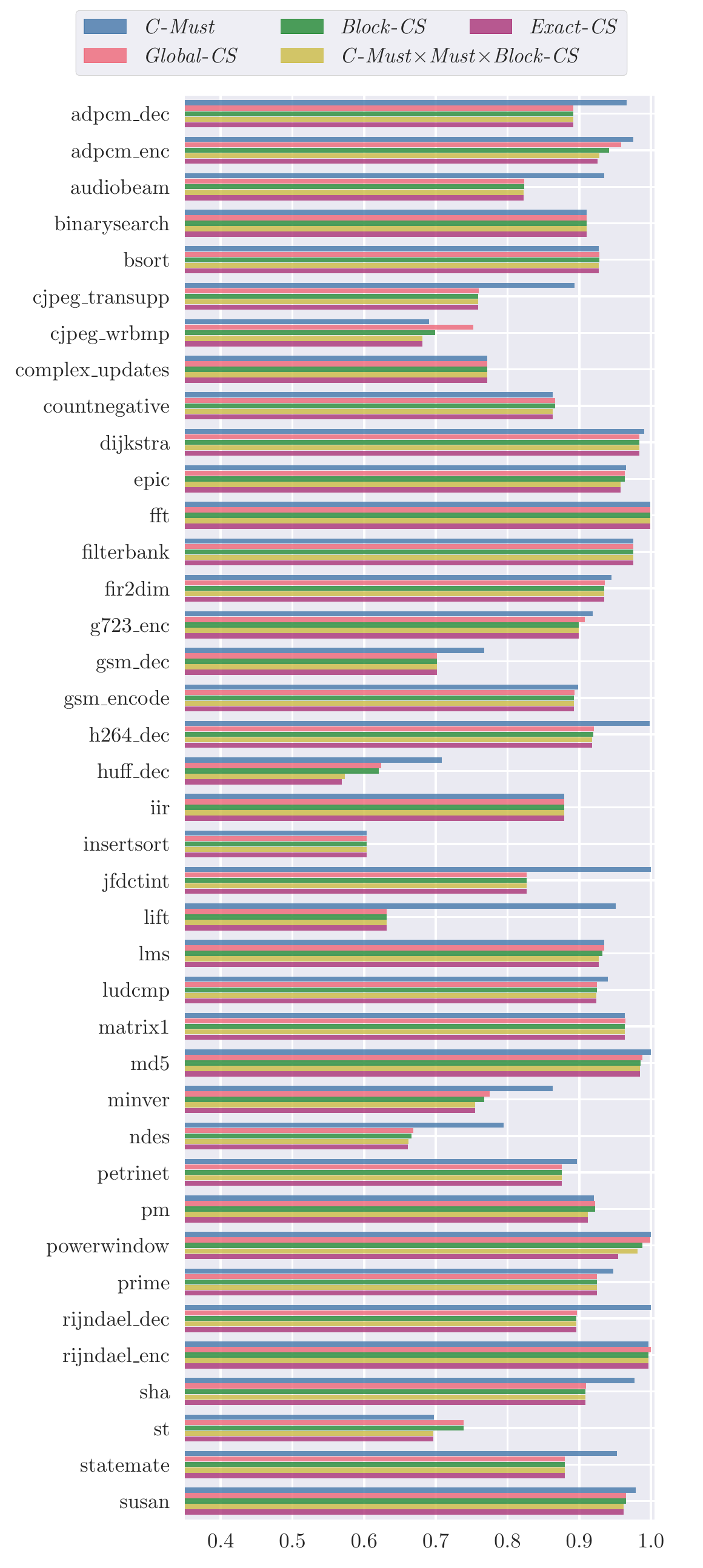}}
            \subcaption{\emph{Without} compiler optimizations.}
        \end{subfigure}
        \begin{subfigure}[c]{0.5\textwidth}
            \centering%
            \scalebox{0.8}{\includegraphics{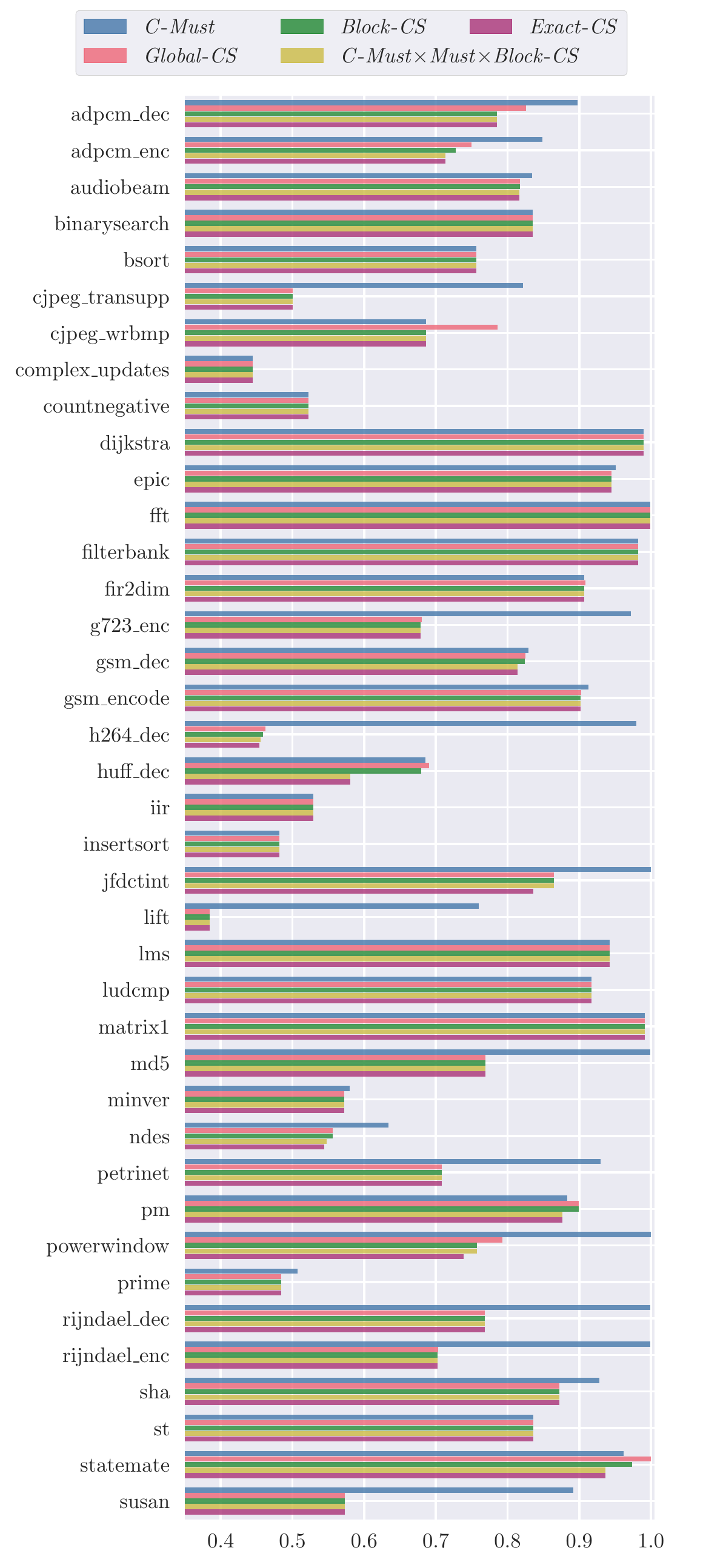}}
            \subcaption{\emph{With} compiler optimizations.}
        \end{subfigure}
        \caption{WCET ratios of the persistence analyses compared with performing no persistence analysis at all.\\Cache configuration: $32$ cache sets, $8$ ways, $16\,\mathrm{B}$ line size, \emph{higher latency of 100~cycles to deliver the first word}.}
    \end{figure*}

    \begin{figure*}[p]
        \begin{subfigure}[c]{0.5\textwidth}
            \centering%
            \scalebox{0.8}{\includegraphics{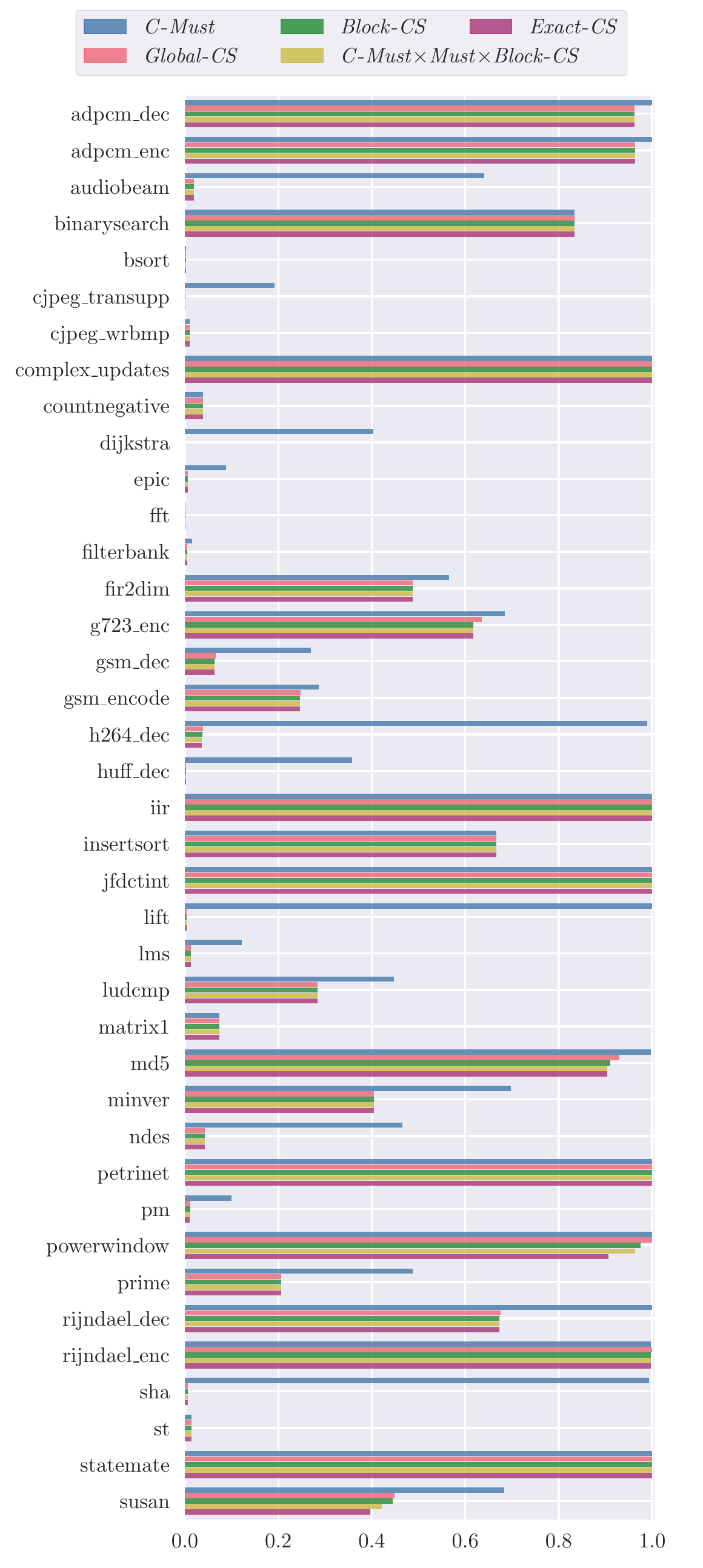}}
            \subcaption{\emph{Instruction} cache misses.}
        \end{subfigure}
        \begin{subfigure}[c]{0.5\textwidth}
            \centering%
            \scalebox{0.8}{\includegraphics{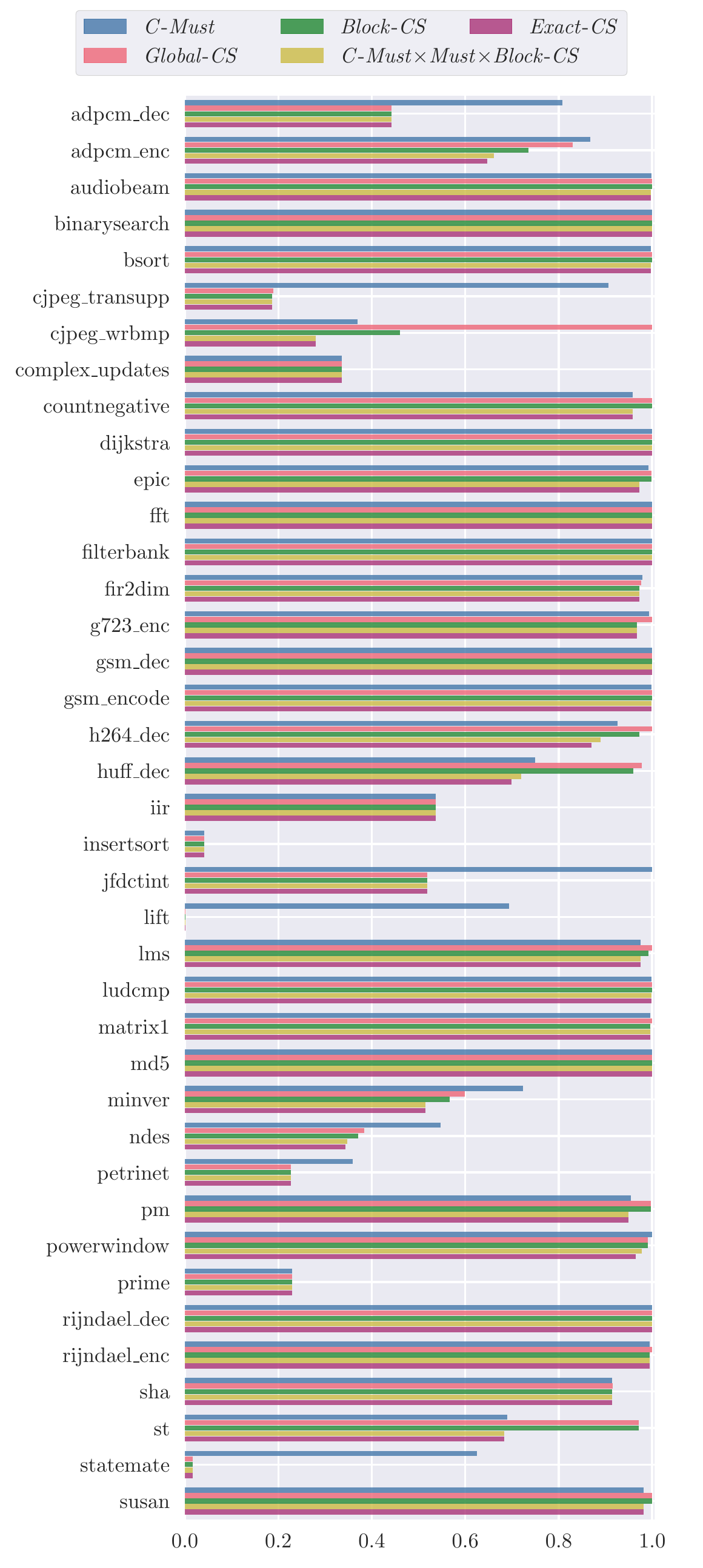}}
            \subcaption{\emph{Data} cache misses.}
        \end{subfigure}
        \caption{Ratios of the maximum number of instruction and data cache misses of the persistence analyses compared with performing no persistence analysis at all.\\Cache configuration: $32$ cache sets, $8$ ways, $16\,\mathrm{B}$ line size, \emph{without} compiler optimizations.}
    \end{figure*}

    \begin{figure*}[p]
        \begin{subfigure}[c]{0.5\textwidth}
            \centering%
            \scalebox{0.8}{\includegraphics{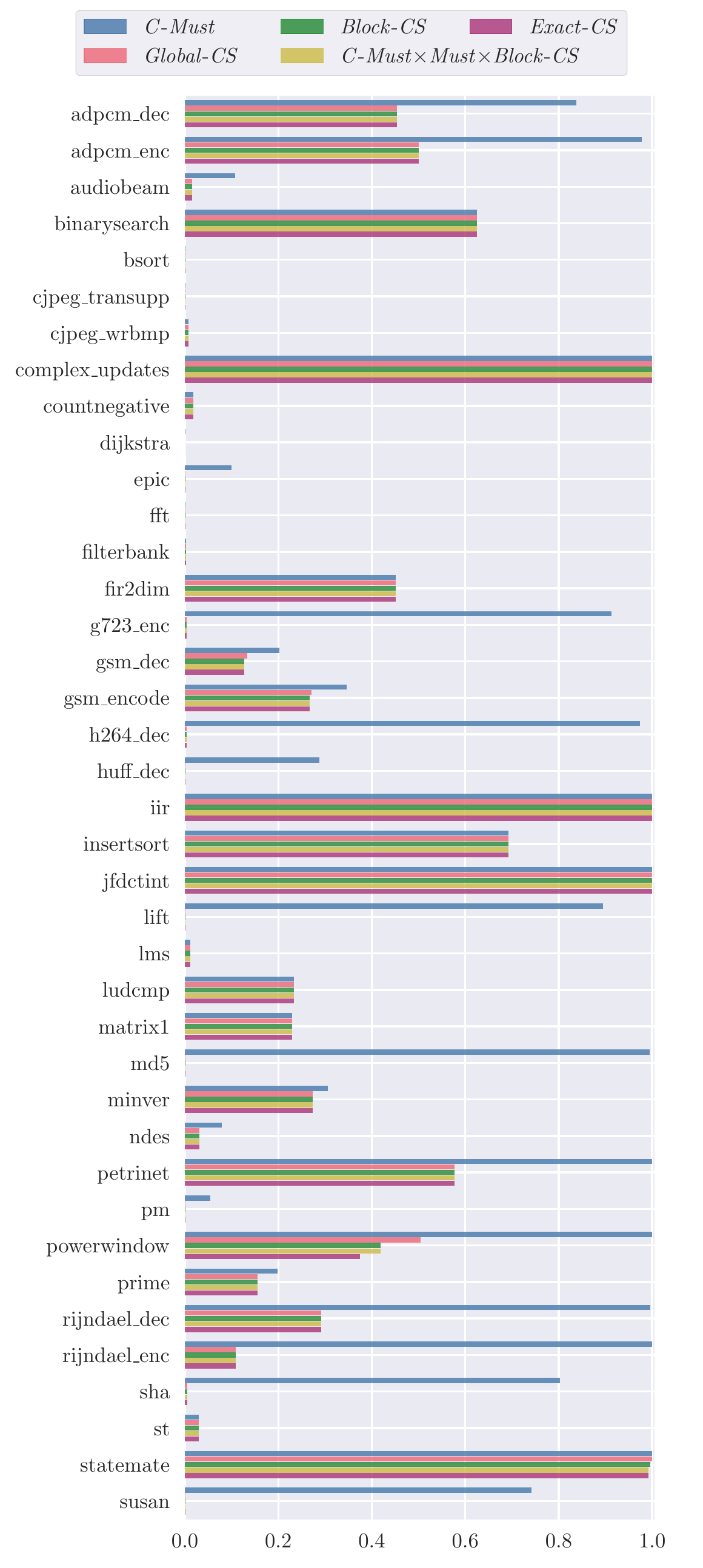}}
            \subcaption{\emph{Instruction} cache misses.}
        \end{subfigure}
        \begin{subfigure}[c]{0.5\textwidth}
            \centering%
            \scalebox{0.8}{\includegraphics{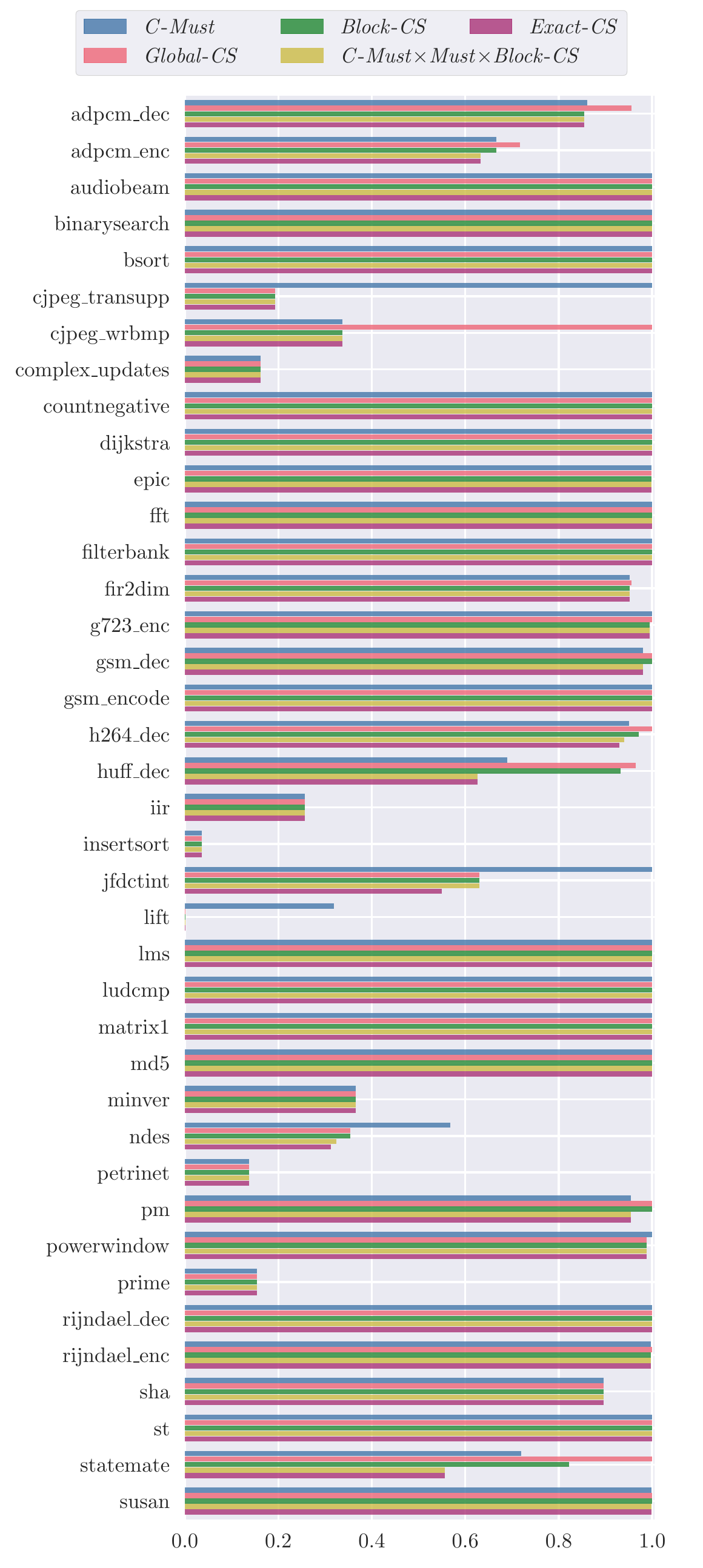}}
            \subcaption{\emph{Data} cache misses.}
        \end{subfigure}
        \caption{Ratios of the maximum number of instruction and data cache misses of the persistence analyses compared with performing no persistence analysis at all.\\Cache configuration: $32$ cache sets, $8$ ways, $16\,\mathrm{B}$ line size, \emph{with} compiler optimizations.}
    \end{figure*}

    \begin{figure*}[p]
        \begin{subfigure}[c]{0.5\textwidth}
            \centering%
            \scalebox{0.8}{\includegraphics{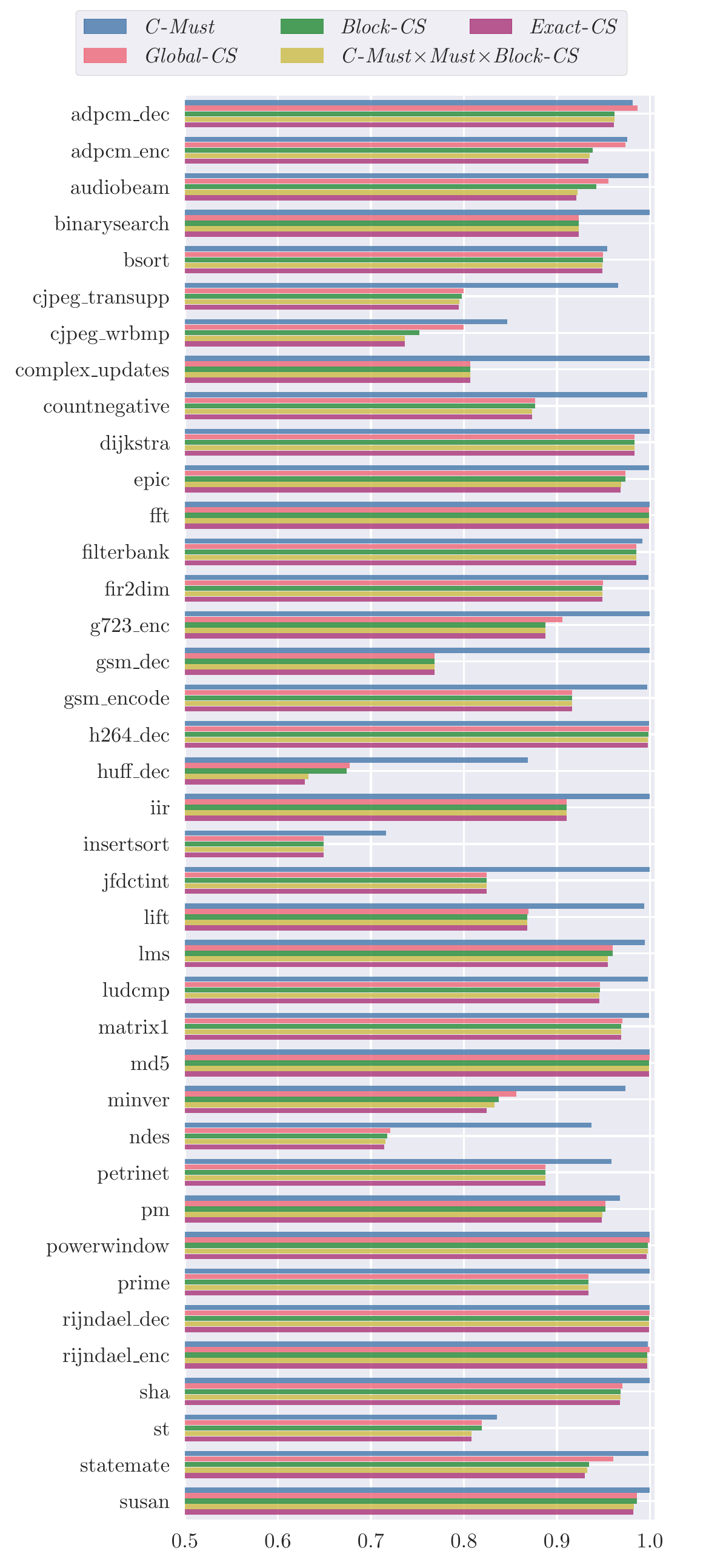}}
            \subcaption{\emph{Without} compiler optimizations.}
        \end{subfigure}
        \begin{subfigure}[c]{0.5\textwidth}
            \centering%
            \scalebox{0.8}{\includegraphics{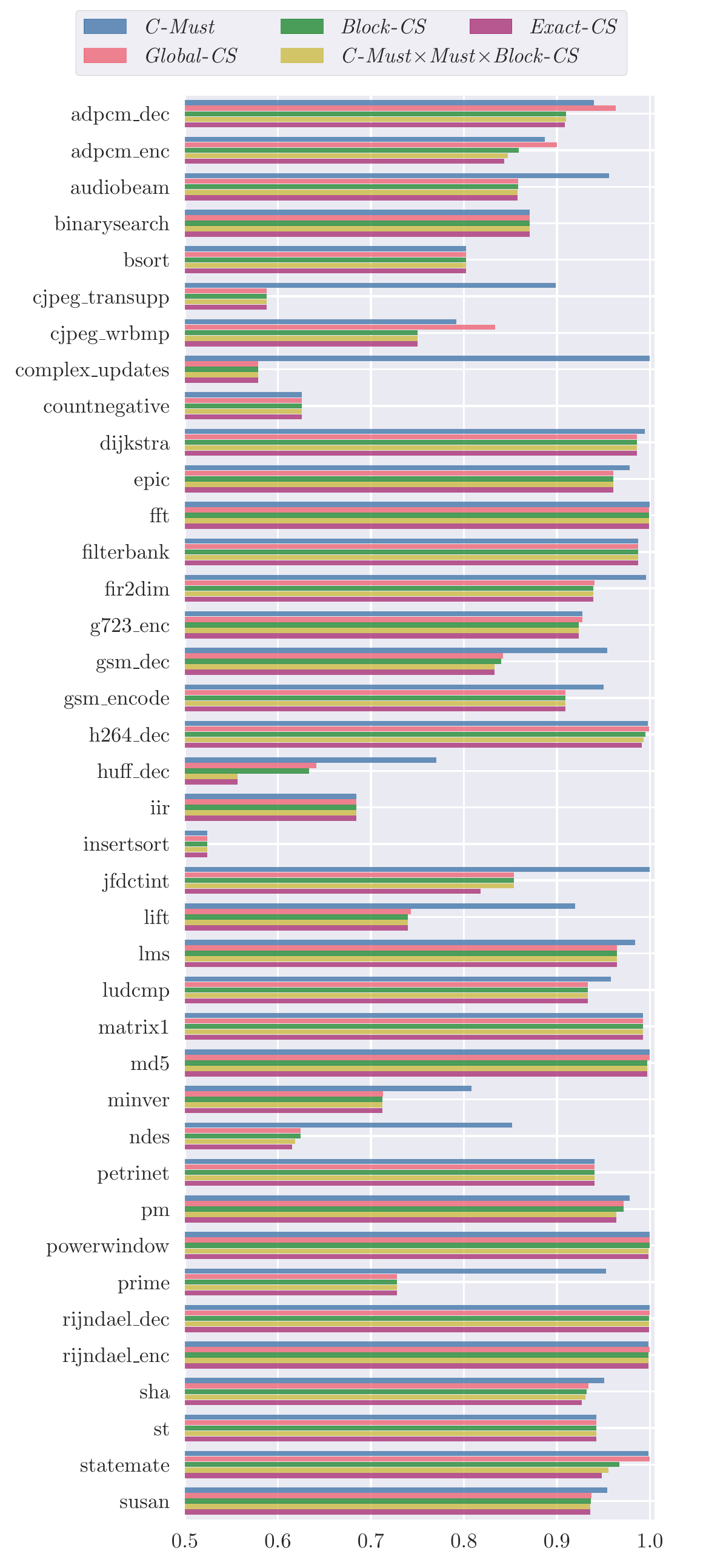}}
            \subcaption{\emph{With} compiler optimizations.}
        \end{subfigure}
        \caption{WCET ratios of the persistence analyses compared with performing no persistence analysis at all.\\Cache configuration: \emph{$8$ cache sets}, $8$ ways, $16\,\mathrm{B}$ line size.}
    \end{figure*}

    \begin{figure*}[p]
        \centering%
        \begin{subfigure}[b]{.5\linewidth}
            \centering
            \scalebox{0.75}{\input{scatter-time-8-32.pgf}}
            \caption{Time, \emph{without} compiler optimizations.}
        \end{subfigure}%
        \begin{subfigure}[b]{.5\linewidth}
            \centering
            \scalebox{0.75}{\input{scatter-time-8-32-opt.pgf}}
            \caption{Time, \emph{with} compiler optimizations.}
        \end{subfigure}%

        \vspace{1cm}

        \begin{subfigure}[b]{.5\linewidth}
            \hspace{5pt}
            \centering
            \scalebox{0.75}{\input{scatter-mem-8-32.pgf}}
            \caption{Memory, \emph{without} compiler optimizations.}
        \end{subfigure}%
        \begin{subfigure}[b]{.5\linewidth}
            \hspace{5pt}
            \centering
            \scalebox{0.75}{\input{scatter-mem-8-32-opt.pgf}}
            \caption{Memory, \emph{with} compiler optimizations.}
        \end{subfigure}%
        \caption{Run time and memory comparison of persistence analyses relative to $\gcs$.\\Cache configuration: $32$ cache sets, $8$ ways, $16\,\mathrm{B}$ line size.}
    \end{figure*}

    \begin{figure*}[p]
        \centering%
        \begin{subfigure}[b]{.5\linewidth}
            \centering
            \scalebox{0.75}{\input{scatter-time-8-8.pgf}}
            \caption{Time, \emph{without} compiler optimizations.}
        \end{subfigure}%
        \begin{subfigure}[b]{.5\linewidth}
            \centering
            \scalebox{0.75}{\input{scatter-time-8-8-opt.pgf}}
            \caption{Time, \emph{with} compiler optimizations.}
        \end{subfigure}%

        \vspace{1cm}

        \begin{subfigure}[b]{.5\linewidth}
            \hspace{5pt}
            \centering
            \scalebox{0.75}{\input{scatter-mem-8-8.pgf}}
            \caption{Memory, \emph{without} compiler optimizations.}
        \end{subfigure}%
        \begin{subfigure}[b]{.5\linewidth}
            \hspace{5pt}
            \centering
            \scalebox{0.75}{\input{scatter-mem-8-8-opt.pgf}}
            \caption{Memory, \emph{with} compiler optimizations.}
        \end{subfigure}%
        \caption{Run time and memory comparison of persistence analyses relative to $\gcs$.\\Cache configuration: \emph{$8$ cache sets}, $8$ ways, $16\,\mathrm{B}$ line size.}
    \end{figure*}

    \clearpage 
\end{techreport}

\bibliographystyle{IEEEtran}
\balance
\bibliography{expers-tech-report}

\end{document}